\documentclass[twocolumn]{aastex631}

\newcommand{\Add}[1]{#1}
\newcommand{\AddSec}[1]{#1}

\newcommand{\AGAMA}{\texttt{AGAMA} }
\newcommand{\ASURA}{\texttt{ASURA} }

\shorttitle{AASTeX v6.3.1 Sample article}
\shortauthors{Yutani et al.}
\graphicspath{{./}{./}}
\begin{document}

%\title{Rapid Clumpy Accretion via Damped Orbits in isolated spiral galaxy}
\title{Dynamical Mechanism of Episodic Gas Accretion to the Central Region of Spiral Galaxies}

\author[0009-0008-8432-7460]{Naomichi Yutani}
\affiliation{Department of Planetology, Graduate School of Science, Kobe University 1-1 Rokkodai, Nada-ku, Kobe, Hyogo 657-8501, Japan}
\email{yutaninm@gmail.com}

\author[0000-0003-2535-5513]{Nozomu Kawakatu}
\affiliation{National Institute of Technology, Kure College, 2-2-11, Agaminami, Kure, Hiroshima, 737-8506, Japan}

\author[0000-0001-8226-4592]{Takayuki Saitoh}
\affiliation{Department of Planetology, Graduate School of Science, Kobe University 1-1 Rokkodai, Nada-ku, Kobe, Hyogo 657-8501, Japan}
\affiliation{Center for Planetary Science (CPS), Graduate School of Science, Kobe University 1-1Rokkodai, Nada-ku, Kobe, Hyogo 657-8501, Japan}

\author[0000-0002-8779-8486]{Keiichi Wada}
\affiliation{Kagoshima University, Graduate School of Science and Engineering, Kagoshima 890-0065, Japan}
\affiliation{Research Center for Space and Cosmic Evolution, Ehime University, 2-5 Bunkyo-cho, Matsuyama, Ehime 790-8577, Japan}
\affiliation{Hokkaido University, Faculty of Science, Sapporo 060-0810, Japan}

%% Note that the \and command from previous versions of AASTeX is now
%% depreciated in this version as it is no longer necessary. AASTeX 
%% automatically takes care of all commas and "and"s between authors names.

%% AASTeX 6.31 has the new \collaboration and \nocollaboration commands to
%% provide the collaboration status of a group of authors. These commands 
%% can be used either before or after the list of corresponding authors. The
%% argument for \collaboration is the collaboration identifier. Authors are
%% encouraged to surround collaboration identifiers with ()s. The 
%% \nocollaboration command takes no argument and exists to indicate that
%% the nearby authors are not part of surrounding collaborations.

%% Mark off the abstract in the ``abstract'' environment. 
\begin{abstract}
We performed \textit{N}-body/SPH simulations of isolated spiral galaxies with various bulge-to-disk mass ratios ($M_{\rm bulge}/M_{\rm disk}$) from 0.02 to 0.2 to investigate mass transport from galactic scales (10 kpc) down to circumnuclear disk scales ($\lesssim$ 100 pc).
Our analysis revealed these main findings,
(1) Gravitational torque from stellar spiral arms causes gas accretion with $\sim1\,M_\odot\,{\rm yr}^{-1}$  along the gas spiral arms from a few kpc to a few 100 pc scale.
The density of accreting gas is a few 100 ${\rm cm^{-3}}$, comparable to the gas arms. 
The pressure gradient force is over an order of magnitude weaker than the stellar gravitational torque.
(2) Gravitational torque from barred structure causes episodic gas clump accretion with $\sim1\,M_\odot\,{\rm yr}^{-1}$ on timescales of 10 Myr from kpc to a few 100 pc scale.
The densities of these clumps exceed 700 ${\rm cm^{-3}}$, and this accretion occurs along elliptical orbits with a delayed phase relative to the bar potential \citep{wada1994}.
(3) Episodic gas clumpy accretion is important for galactic center instability, confirmed by $M_{\rm bulge}/M_{\rm disk}$ = 0.02 but not by $M_{\rm bulge}/M_{\rm disk}$ = 0.1 and 0.2. This difference occurs because in the bulge-dominated potentials, bar instability is suppressed and rapid gas clumps accretion does not occur.
These findings suggest that gas clump accretion events driven by bars could be a source of high-density gas to the galactic center of the spiral galaxy, potentially promoting temporary activity in the galactic center.
\end{abstract}

%% Keywords should appear after the \end{abstract} command. 
%% The AAS Journals now uses Unified Astronomy Thesaurus concepts:
%% https://astrothesaurus.org
%% You will be asked to selected these concepts during the submission process
%% but this old "keyword" functionality is maintained in case authors want
%% to include these concepts in their preprints.

\section{INTRODUCTION} \label{sec:intro}
% The relationship between SMBH growth and galaxy evolution
In recent decades, a growing body of observational and theoretical evidence has revealed that supermassive black holes (SMBHs), residing at the centers of galaxies, evolve with their host galaxy. 
This is indicated by the empirical scaling relations between SMBH mass and galactic bulge properties, such as stellar mass and velocity dispersion \citep[e.g.,][]{marconi2003,burkert2010,kormendy2013,mcconnell2013}. 
These correlations suggest that the processes governing galaxy evolution - such as star formation, gas accretion, and dynamical evolution of galaxies - are intimately linked to the mechanisms that drive SMBH growth \citep{burkert2001}.
In this context, understanding how SMBHs obtain their mass is not only essential for black hole physics but also a key component in unraveling the broader picture of galaxy evolution across cosmic time.\par

% The relationship between dense gas in galactic center and SMBH growth
Recent observational efforts have focused on resolving the dense molecular gas in galactic nuclei to reveal the mechanisms of SMBH fueling.
The dense gas of the circumnuclear disk (CND), typically traced by high-density tracers such as HCN or HCO$^{+}$, is now recognized as necessary for maintaining or triggering SMBH accretion \citep{izumi2016}. High-resolution ($\sim$ sub-pc) observations, particularly those from ALMA, have begun to reveal the complexity of the circumnuclear environment \citep{izumi2023}. 
Studies have reported the presence of CNDs, nuclear bars, nuclear spirals, and gas streamers that trace the flow of material from kiloparsec scales toward the central few hundred parsecs \citep[e.g.,][]{burillo2014,izumi2018,combes2013,combes2019}.
However, a crucial open question remains: how is this gas transported inward from galactic scales to the circumnuclear region where it can actively feed the SMBH?

% The accretion mechanism and clumpy formation
Various theoretical models have proposed mechanisms through which gas can lose angular momentum and be transported from galactic scales to the central regions of galaxies.
In particular, gravitational torques induced by galaxy mergers \citep[e.g.,][]{mihos1996,hopkins2010,kawaguchi2020} and violent disk instabilities in gas-rich systems \citep[e.g.,][]{noguchi1999,dekel2009,bournaud2011} are recognized as the two primary drivers of large-scale inflow.
These dynamical processes not only facilitate angular momentum redistribution but also trigger the gravitational fragmentation of unstable gas disks, naturally leading to the formation of massive ($\sim10^7-10^9 M_\odot$) gas clumps \citep{bournaud2009,matsui2012,renaud2015,inoue2016,nakazato2024}. 

After their formation, these giant clumps are expected to interact gravitationally with their host disks and with each other, losing angular momentum through torques and dynamical friction. 
As a result, they may gradually migrate toward the galactic center, delivering large amounts of dense gas into the circumnuclear region and potentially fueling both bulge growth and black hole accretion. Although the survival of gas clumps is debatable \citep[e.g.,][]{bournaud2012, trump2014}, understanding the physical processes of this clump-driven mass transport is thus crucial to unveiling the mechanisms that couple star formation, black hole fueling, and galaxy evolution.\par

\Add{\cite{goldbaum2015, goldbaum2016} showed that gravitational instability can maintain Toomre $Q$ values at around unity even in the absence of stellar feedback. In addition, they found that stellar feedback locally suppresses the formation of massive star-forming clouds. These results suggest that, while feedback strongly influences the formation and survival of gas clouds, the galactic potential is expected to have a strong impact on their dynamics.}\par

%研究内容
In this study, to clarify the accretion mechanism in galaxies, isolated gas-rich disk galaxy simulations were performed \Add{with different mass ratios of bulge to disk, because the spherical potential is controls mass transport processes in galactic disk \citep{ostriker1973,athanassoula2003}.}
In particular, we analyzed the torque on high-density gas, tracked these gas clumps, and evaluated their contribution to the mass supply to the galactic center for a few 100 Myr.\par

%章立てと宇宙論パラメータ
The remainder of this paper is organized as follows. Section \ref{sec:method} describes the simulation methods and models; Section \ref{sec:galaxy-mor} summarizes the relationship between galaxy morphology and gravitational potential.
Section \ref{sec:spiral-accretion} reports the steady mass accretion mechanism by the spiral arms and the mass accretion rate at each radius.
In Section \ref{sec:clumpy-acc}, we summarize the clumpy mass accretion mechanism by bar potentials.
Section \ref{sec:discuss} discusses the lifetime of gas clumps and the conditions of clumpy accretion via damped orbit.\par

\begin{figure*}
	%\epsscale{1.1}
	%\plotone{Yutani25-mod.pdf}
	\includegraphics[width=\textwidth]{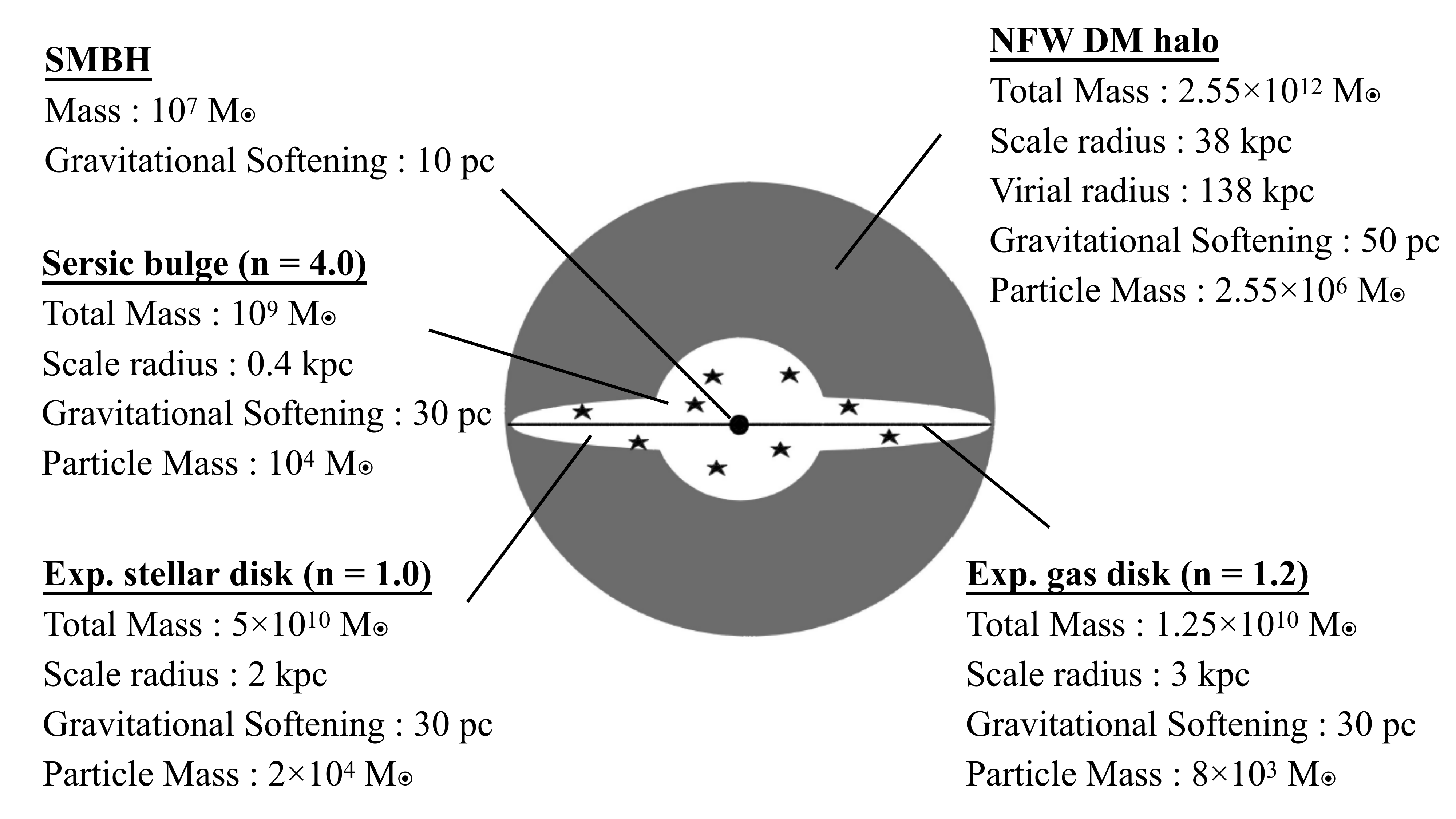}
	\caption{Initial setup of the fiducial model.\label{fig:yutani25-fidu}}
\end{figure*}

\section{Models and Methods}\label{sec:method}
We simulated an isolated disk galaxy with various \Add{bulge-to-disk mass ratios} using \Add{an} \textit{N}-body/smoothed particle hydrodynamics (SPH) simulation code \texttt{ASURA} \citep{saitoh2008, saitoh2013}.
Since we are particularly interested in the accretion mechanism, we used the particle-based method, which allows us to directly trace the accretion process of the gas clumps.

\subsection{Numerical Methods}\label{sec:numerical}
We solve the self-gravitational interactions of the gas, stellar, dark matter (DM), and BH particles to assess the self-gravitational instability of a disk galaxy. 
We adopt a parallel tree method following the parallel strategy of \cite{makino2004}. 
To accelerate the computation of the gravitational interactions, we use \texttt{Phantom-GRAPE} \citep{tanikawa2013}, which is a software emulator of \texttt{GRAPE} \citep{sugimoto1990}.

The dynamics of \Add{the} interstellar medium is calculated with Density-Independent SPH \citep{saitoh2013}, a method that can correctly handle contact discontinuities.
Hydrodynamics and radiative cooling are treated with \Add{an} operator splitting \Add{method}: the hydrodynamic variables are advanced explicitly, and the subsequent cooling sub-step is integrated with the exact \Add{cooling} solver of \cite{zhu2017}.
The cooling function was calculated using $\tt Cloudy$ \citep{ferland1998} under the optically thin assumption based on gas metallicity.
We use ${\tt CELib}$ \citep{saitoh2017} to \Add{handle} the chemical evolution of star particles.
The initial gas metallicity was set to one solar metallicity.

\subsection{The Treatment of Star Formation}
\Add{Star formation is reproduced in a stochastic manner by converting a part of gas particles into starparticles \citep{saitoh2008, saitoh2009}.
Once a gas particle satisfies the following three conditions, (1) $n_{\rm H}\ >\ 100\ \mathrm{cm}^{-3}$, (2) $T\ <\ 1000\ \mathrm{K}$, and (3) $\nabla \cdot \boldmath{v}\ <\ 0$, a star particle is spawned from the gas particleunder the Schmidt law \citep{schmidt1959}:
\begin{equation}
	\frac{d \rho_*}{dt} = \epsilon_{\rm SF} \frac{\rho_{\rm gas}}{t_{\rm dyn}},\label{eq:sf}
\end{equation}
where $\rho_{*}$ and $\rho_{\rm gas}$ are the stellar and gas densities, respectively, and $\epsilon_{\rm SF}$ is the local star formation efficiency, and $t_{\rm dyn}$ is the local dynamical time of the gas. 
Here, we assume 0.033 for the local star formation efficiency. 
The stellar mass is set to one-third of the original gas mass and the whole gas is converted into a single star particle when the gas mass is less than one-third of the original gas mass. 
This model can reproduce the Schimidt-Keniccutt relation \citep{kennicutt1998}. 
A simple stellar population (SSP) approximation is adopted for the star particles. Thus, a star particle consists of a group of stars sharing their formation time and metallicities that take over from the parent gas particles. We adopt the Salpeter initial mass function (IMF) with a mass range of 0.1 to 120 {$\rm M_\odot$.}}\par

\subsection{The Treatment of Stellar Feedback}
\Add{In our simulations, we implemented a momentum feedback model of Type II SNe (supernovae) following the FIRE-2 model \citep{hopkins2018}. 
We count individual SNe in each SSP particle and handle their explosion events discretely.
We assume that each SN releases $\eta_{\rm SN} \times 10^{51}$ ergs of energy for the surrounding ISM.
We adopt $\eta_{\rm SN} = 1$ as the fiducial model, but we investigate the impact of varying $\eta$, particularly on the survival of gas clumps, in Appendix \ref{app:sn-eff}.
When a Type-II SN takes place, the star particle deposits momentum and thermal energy into the surrounding gas particles. 
Since the Sedov-Taylor phase is not resolved in our simulations, the SN feedback is implemented by imparting the radial terminal momentum, $p_{\rm t}$, to the surrounding gas particles, weighted by the SPH kernel. The terminal momentum of the swept-up gas in the momentum-conserving phase from a single explosion can be estimated as
\begin{equation}
	\frac{p_{\rm t}}{\rm M_\odot\ km s^{-1}} \sim 4.8 \times 10^5 \bigg(\frac{E_{\rm ej}}{10^{51} {\rm erg}}\bigg)^{\frac{13}{14}}\bigg(\frac{n}{\rm cm^{-3}}\bigg)^{-\frac{1}{7}}\bigg(\frac{Z}{Z_\odot}\bigg)^{-0.21},\label{eq:sn}
\end{equation}
where $n$ and $Z$ are the number density and metallicity of the ambient gas, respectively, as shown in \cite{hopkins2018}. 
If the gas number density is lower than 0.01 ${\rm cm^{-3}}$, the ejecta energy $E_{\rm ej}$ is instead distributed as thermal energy to the surrounding gas particles, weighted by the SPH kernel.}

\Add{Along with the energy feedback, chemical evolution is also solved simultaneously. Chemical evolution of eighteen elements is solved using a chemical evolution library ${\tt CELib}$ \citep{saitoh2017}. We adopt the yield table of \cite{nomoto2013}. The released elements are redistributed to the surrounding gas particles with the weights evaluated by the SPH kernel.}\par

\AddSec{Assuming a Salpeter IMF and using the \cite{nomoto2013} yields, the total energy of Type II SNe is 3.56 $\times$ 10$^{48}$ erg per solar mass. For a star particle with 2 $\times$ 10$^4$ $M_{\rm \odot}$, this amounts to 7.12 $\times$ 10$^{51}$ erg, corresponding to 7 Type II SNe events.}\par

\AddSec{When a star particle is formed, we integrate the Salpeter IMF to determine how many massive stars it contains that will trigger an SNe event. For each such star, we assign an explosion time using stellar lifetime tables and then trigger a single supernova accordingly. After one event is triggered, we repeat the procedure for the next lower mass range until no further massive stars remain. }\par

\Add{In addition, thermal feedback from HII regions is taken into account by heating the gas particles surrounding young stars to 10$^4$ K at each time step. The heating region is determined by a St$\rm \ddot{r}$omgren volume approach \citep{hopkins2012, baba2015}. The time-evolving metallicity dependent ionizing photon number released from each star particle is calculated using a stellar population synthesis code ${\tt PEGASE}$ \citep{fioc1999}.}

\subsection{Galaxy Models}
Figure \ref{fig:yutani25-fidu} shows a schematic of the fiducial model.
We employed a fully self-consistent \textit{N}-body simulation to represent not only the DM halo, but also the bulge and disk components of galaxies to handle dynamical friction or disk instability.
The self-consistent initial conditions for the mass distribution and velocity of each particle are obtained using \AGAMA \citep{vasiliev2019}. 
The \AGAMA library is a set of tools for building and analyzing models of galaxies, including the generation of self-consistent initial conditions for multiple components consisting of disks and spherical systems, such as Figure \ref{fig:yutani25-fidu}.\par

\begin{figure}
	%\epsscale{0.8}
	%\plotone{Yutani25-vc.png}
	\includegraphics[width=0.45\textwidth]{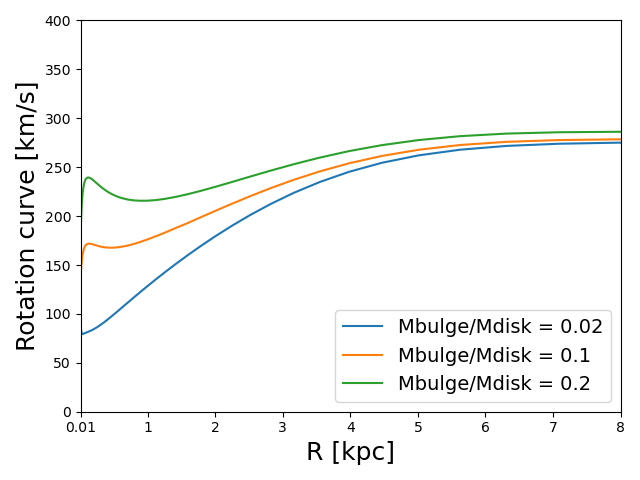}
	\caption{Comparison of rotation curve in different \Add{potentials}. Blue line is fiducial model, orange line is classical bulge model, green line is heavy bulge model.}\label{fig:yutani25-vc}
\end{figure}

% Write how to choose phisical parameter of fiducial model
The fiducial model represents a typical galaxy, which includes \Add{an} SMBH, DM halo, galactic bulge\Add{,} and disks of gas and stars. 
We set the initial SMBH mass to $10^{7}\,M_\odot$ and adopt a stellar-to-halo mass ratio of 2$\%$. 
This ratio corresponds to galaxies whose dark-matter halos have $M_{\rm halo}\!\simeq\!10^{12}\,M_\odot$, consistent with the observed stellar-halo mass relation at $z\!\sim\!2$ \citep[e.g.,][]{matthee2017,behroozi2019}.
In our simulations, the virial mass of the NFW DM halo is $M_{\rm halo}\!\approx\!3\times10^{12}\,M_\odot$; their virial radii and scale radii are calculated with \texttt{COLOSSUS} \citep{diemer2018} using the typical $z\!\sim\!2$ cosmology.

% Write other models, and why choose bulge mass as free parameter.
In order to investigate the relationship between bulge mass and mass accretion mechanism, we calculated \Add{three bulge to stellar disk mass ratios, 0.02 (fiducial), 0.1, and 0.2,} which together span the typical range seen in observations of local bulges \citep[e.g.,][]{kormendy2013}.
The rotation curves are shown in Figure \ref{fig:yutani25-vc}.
The difference in bulge mass makes a significant change in the potential within the inner kpc of the galaxy.\par

% calculation parameter
The stellar and gas components share the same mass resolution in every run.
For the dark-matter halo, we always use $10^{6}$ particles; consequently the DM particle mass varies from $1.25\times10^{6}$ to $2.75\times10^{6}\,M_\odot$ across the three models.

\begin{figure}
	%\epsscale{1.25}
	%\centering
	%\plotone{Yutani25-gas-star-12kpc.pdf}
	\includegraphics[width=0.45\textwidth]{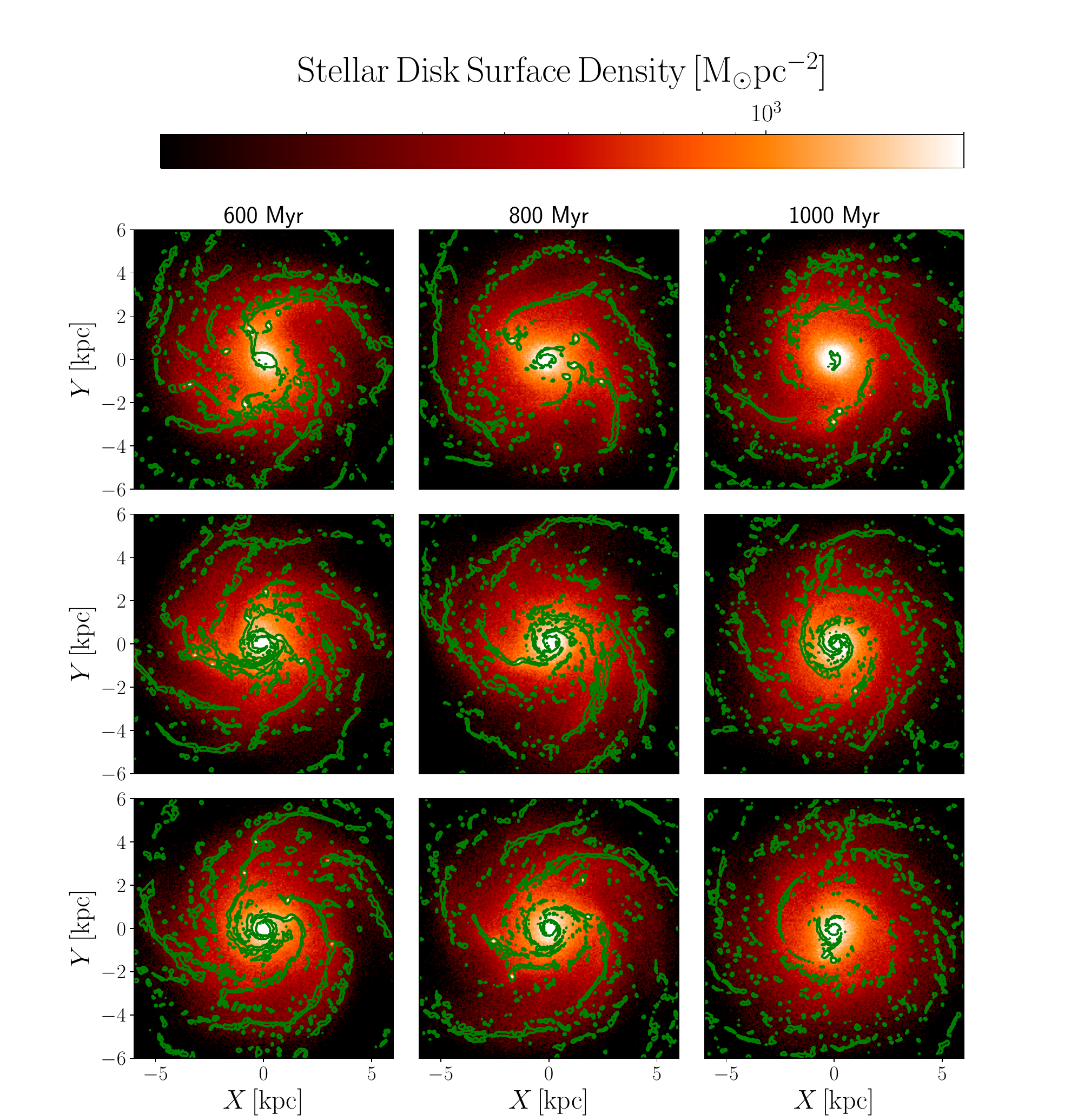}
	\caption{Surface density evolution of stellar disk and gas in three models.
		Columns, arranged left-to-right, show snapshots at 600 Myr, 800 Myr, and 1000 Myr; rows, ordered from top to bottom, correspond to \Add{bulge to stellar disk mass ratios of $M_{\mathrm{bulge}}/M_{\mathrm{disk}} = 0.02,\ 0.10$, and $0.20$.}
		Green contour shows the gas surface density at 50 $M_\odot\ {\rm pc^{-2}}$. Each panel shows a 12 kpc $\times$ 12 kpc region around a galactic center. }\label{fig:yutani25-sd-12kpc}
\end{figure}

\section{results}
\subsection{Stellar and Gas structures in three models}\label{sec:galaxy-mor}
In this section, we present the relation between galaxy morphology and galactic potential.
In every simulation, the stellar-gas disk promptly settles into a grand-design pattern of two to three spiral arms that span several kpc (Figure \ref{fig:yutani25-sd-12kpc}).
These arms emerge because the disk's self-gravity dominates the axisymmetric halo potential at $\sim$ a few kpc, lowering the Toomre $\Add{Q}$ parameter toward unity and allowing swing amplification of non-axisymmetric modes \citep{toomre1981,bt2008}.
Once established, the arms act as a gravitational torque engine: they extract angular momentum from the gas and funnel it inward, a process demonstrated analytically by \citet{lyndenbell1972} and confirmed in a wide range of numerical studies \citep[e.g.,][]{sellwood1984,wada2002}.
The net mass inflow driven by these spiral torques in our three bulge-SMBH configurations is quantified in Section \ref{sec:spiral-accretion}.\par

In our fiducial \Add{model,} the disk dominates the inner potential, so the system quickly develops a strong bar that extends to $\sim$ 2 kpc.
In the two bulge-heavier models, the deeper, more centrally concentrated spheroid stabilises the disk against the global $\Add{m}$ = 2 mode.
This bulge mass dependence is consistent with the classical bar instability criterion, where a large bulge (or halo) increases central shear and velocity dispersion and suppresses amplification of the swing that seeds long bars \citep{ostriker1973,athanassoula2003}.\par

\begin{figure}
	%\epsscale{1.25}
	%\centering
	%\plotone{Yutani25-gas-star-2kpc.pdf}
	\includegraphics[width=0.45\textwidth]{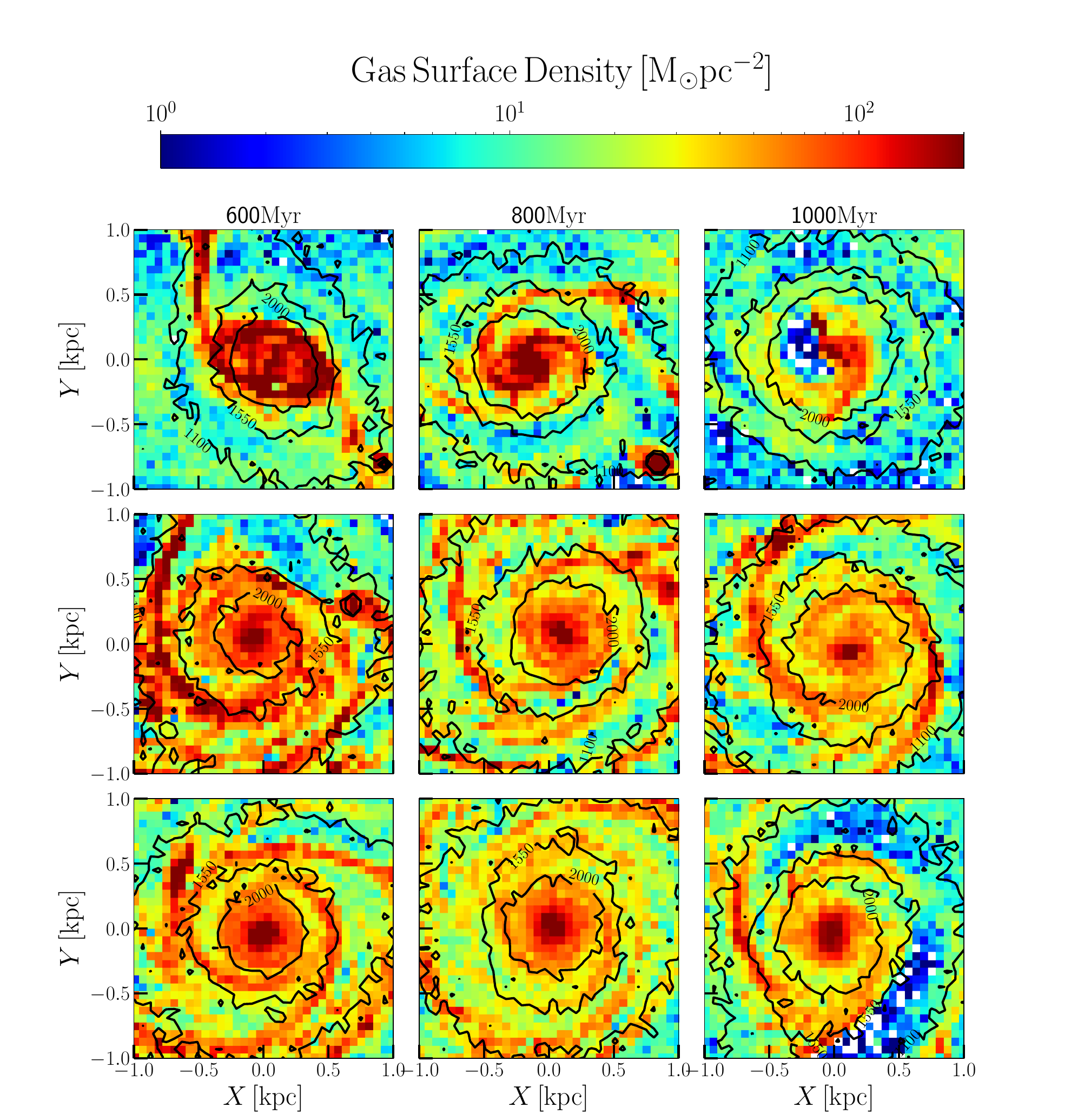}
	\caption{Surface density evolution of gas and stellar disk in three models.
		Columns, arranged left-to-right, show snapshots at 600 Myr, 800 Myr, and 1000 Myr; rows, ordered from top to bottom, correspond to \Add{bulge to stellar disk mass ratios of $M_{\mathrm{bulge}}/M_{\mathrm{disk}} = 0.02,\ 0.10$, and $0.20$.} 
		Contour shows stellar surface density ($M_\odot$ pc$^{-2}$). 
		Each panel shows a 2 kpc $\times$ 2 kpc region around a galactic center. }\label{fig:yutani25-sd-2kpc}
\end{figure}

Self-gravitational instabilities induce clumpy structures.
These clumpy structures are frequent at a few kpc from the galactic center in all models.
Section \ref{sec:clumpy-acc} focuses on the accretion of clumps.\par

In every simulation, the gas-rich disk undergoes local Toomre instabilities and fragments into massive star-forming clumps.
These clumps-typically $10^{7}\,M_\odot$ in mass and a few hundred parsecs across $-$ form most frequently within spiral arms at several kpc radii, independent of the adopted bulge mass.
This gas clump formation in spiral arms closely resembles that of \cite{dobbs2011}.
In our model, small gas clumps are destroyed by \Add{SN} explosions on a timescale of \Add{tens} Myr, while massive gas clumps can survive on a timescale of 100 Myr as discussed in Section \ref{sec:discuss}.
The contribution of these clumps to the overall gas inflow budget is analyzed in Section \ref{sec:clumpy-acc}.\par

%数100 pc まで腕が伸びていることと銀河中心の塊構造に言及する
Figure \ref{fig:yutani25-sd-2kpc} plots the gas and stellar surface densities inside the central kpc.
The kpc-scale spiral arms are connected to a few hundred pc circumferential nuclear disk (CND).
Similar nuclear spirals around CNDs have been reported in ALMA observations of nearby galactic nuclei \citep[e.g.,][]{izumi2018,combes2019,izumi2020}.\par

\begin{figure}
	%\epsscale{1.25}
	%\centering
	%\plotone{Yutani25-gas-mfr.pdf}
	\includegraphics[width=0.45\textwidth]{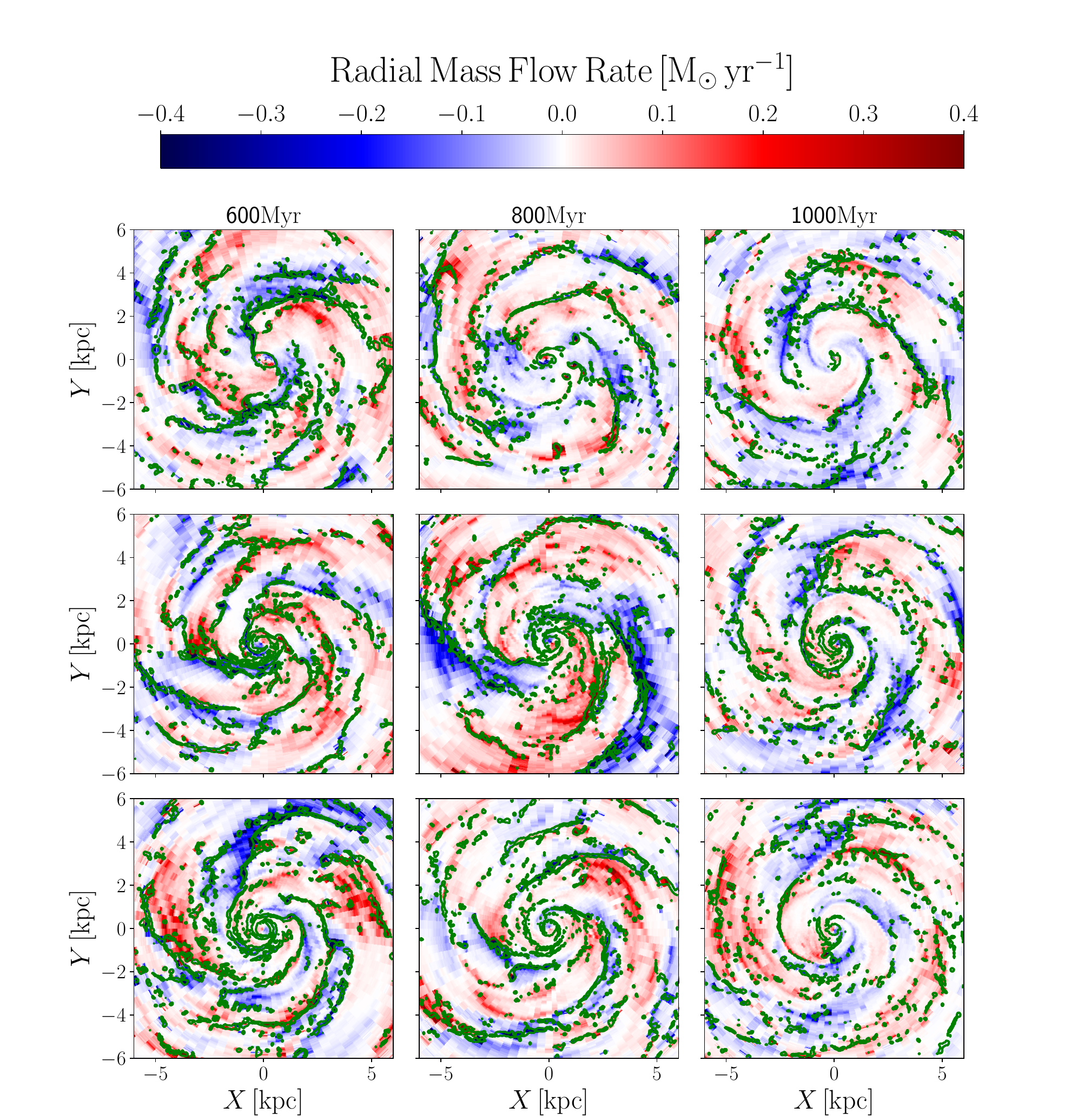}
	\caption{2D radial gas mass flow rate map in three models. 
		Blue region \Add{indicates} inward flow, \Add{while} red region \Add{exhibits} outward flow. 
		Columns, arranged left-to-right, show snapshots at 600 Myr, 800 Myr, and 1000 Myr; rows, ordered from top to bottom, correspond to \Add{bulge to stellar disk mass ratios of $M_{\mathrm{bulge}}/M_{\mathrm{disk}} = 0.02,\ 0.10$, and $0.20$.} 
		Green contour shows the gas surface density at 50 $M_\odot\ {\rm pc^{-2}}$. Each panel shows a 12 kpc $\times$ 12 kpc region around a galactic center.}\label{fig:yutani25-mfr}
\end{figure}

%星の腕が形成する非軸対称構造の中でのガスの運動と自己重力不安定
Within 1 kpc of \Add{the} galactic center, the gas surface density distribution of the fiducial model is significantly different from \Add{the} other two models.
We confirmed \Add{a} nuclear ring or nuclear spiral structure within several 100 pc of the galactic center in the fiducial model, but not in the other two models.
Within 500 pc, the gas distribution is strongly influenced by differences in bulge potential.
In other words, the presence of a large mass bulge (higher rotational speed) contributes to the development of axisymmetric structures by reducing self-gravitational instability \citep{ostriker1973}. \par

Furthermore, Figure \ref{fig:yutani25-sd-2kpc} shows the structure of a clump with surface density greater than 100 \Add{M$_\odot$ pc$^{-2}$}. 
The size of these gas clumps is approximately 100 pc scale. 
This is comparable to the size of giant molecular cloud \citep[e.g.,][]{dessauges2019,chevance2020}.\par

\begin{figure}
	%\epsscale{1.25}
	%\centering
	%\plotone{Yutani25-mfr-r.pdf}
	\includegraphics[width=0.45\textwidth]{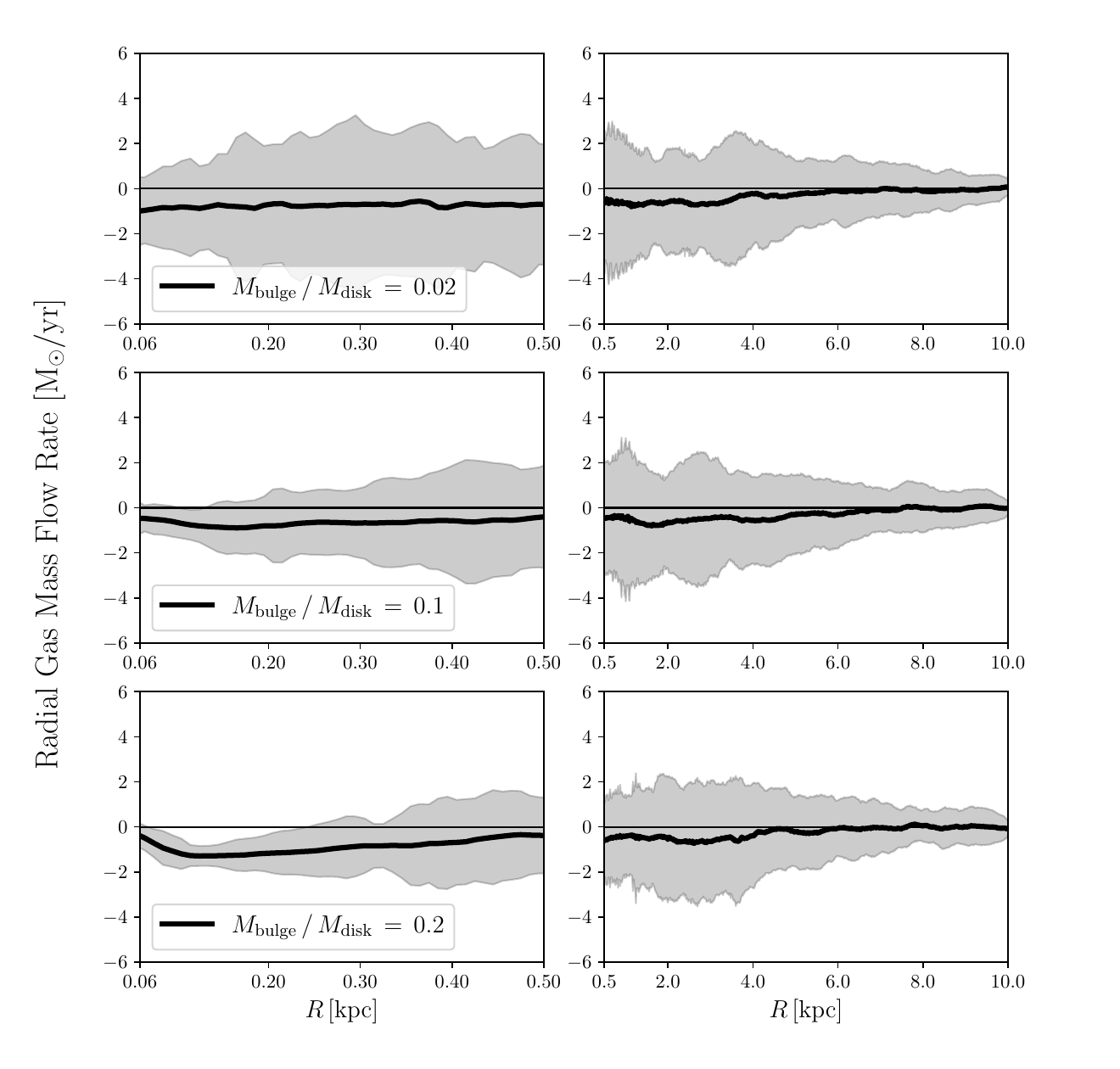}
	\caption{Radial gas mass flow rate in each radius from 60 pc to 10 kpc in three models. The solid black line shows the \Add{time} averaged gas flow rate\Add{,} and the shaded area shows the standard deviation of \Add{the} gas flow rate. To account for the SMBH's motion, mass transport rates from 500 pc to 10 kpc were calculated in a coordinate system stationary with respect to the system's \Add{center} of gravity, whereas the rates from 60 pc to 500 pc were calculated in a coordinate system stationary with respect to the SMBH.}\label{fig:yutani25-mfr-r}
\end{figure}

\subsection{Steady Accretion via Spiral Arms}\label{sec:spiral-accretion}
In this section, we focus on mass transport driven by spiral arms.
From Figure \ref{fig:yutani25-mfr}, we confirm that gas mass transport occurs along the spiral pattern in three models.
Ahead of the gas arm, the radial gas mass flow shows an inward flow (blue regions), whereas between the arms it shows an outward flow (red regions).
Note that this is due to the \Add{epcycle} motion of the gas particles, so it is not the net mass transport rate at each radius.\par

To obtain the net gas mass flow rate, the mass transport rates need to be added together in the azimuthal direction of each radius and averaged over a time scale sufficiently longer than the dynamical time.
Since the dynamical time scale is about 100 Myr for a few kpc radius, we averaged the radial mass flow rate over 400 Myr, from 600 Myr to 1 Gyr.\par

\begin{figure}
	\includegraphics[width=0.45\textwidth]{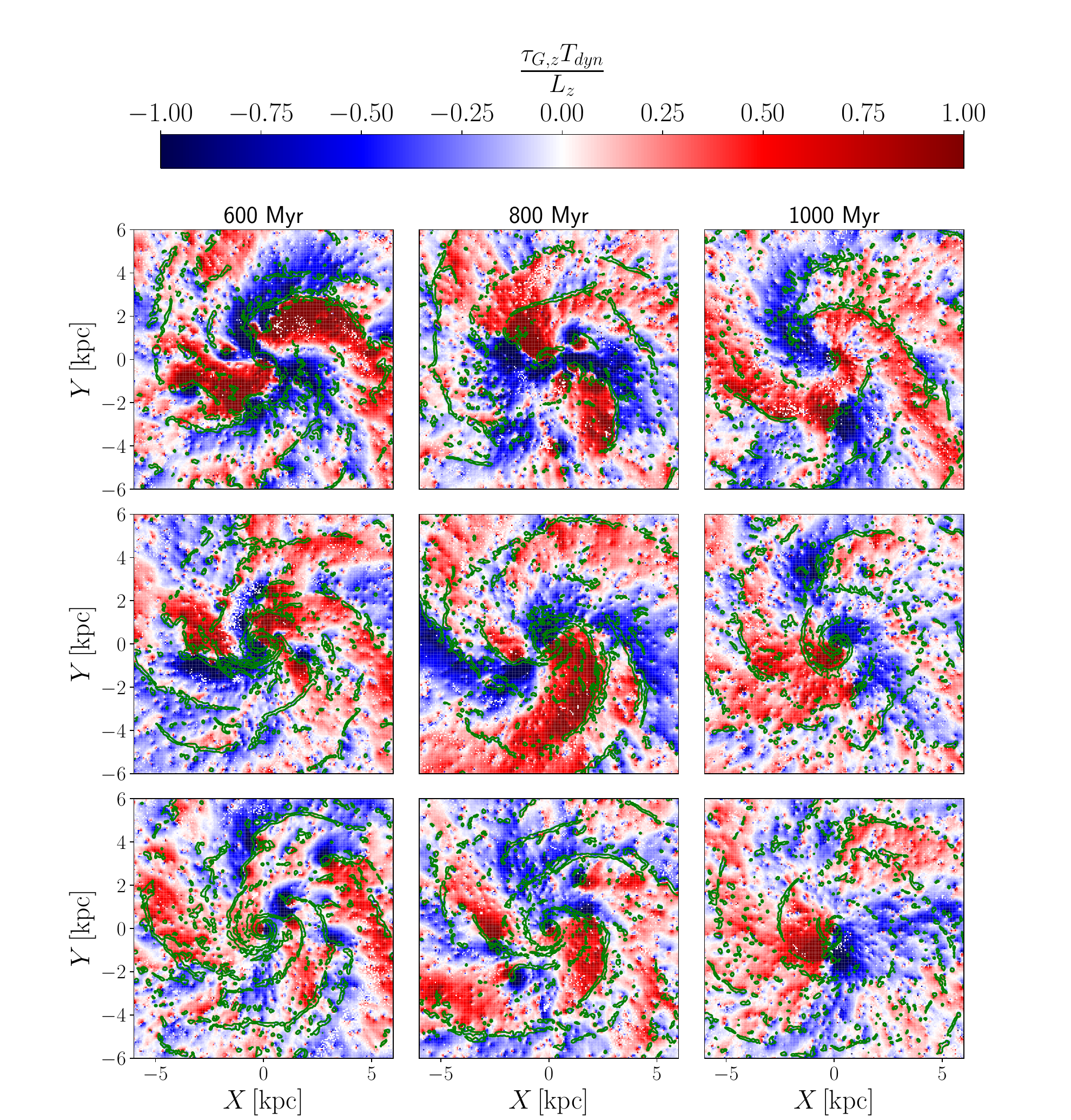}
	\caption{Normalized gravitational torque map in three models. 
		Columns, arranged left-to-right, show snapshots at 600 Myr, 800 Myr, and 1000 Myr; rows, ordered from top to bottom, correspond to \Add{bulge to stellar disk mass ratios of $M_{\mathrm{bulge}}/M_{\mathrm{disk}} = 0.02,\ 0.10$, and $0.20$.} 
		Green contour shows the gas surface density at 50 $M_\odot\ {\rm yr^{-1}}$. Each panel shows a 12 kpc $\times$ 12 kpc region around a galactic center.}\label{fig:yutani25-grav}
\end{figure}

Figure \ref{fig:yutani25-mfr-r} shows that there is 1 $M_\odot\ {\rm yr^{-1}}$ of radial gas mass flow rate from a few 100 pc to a few kpc in all models, indicating net mass accretion along the arms.
However, inside a radius of 500 pc, the dispersion between the massive disk model and the other two models differs significantly.
The dispersion represents the magnitude of the dispersion of radial motion with respect to time. In other words, dispersion is greater in regions where elliptical orbits dominate.
As the dispersion increases when elliptical orbits dominate and approaches zero when circular \Add{orbits} dominate, the gas surface density map in Figure \ref{fig:yutani25-sd-2kpc} and the mass transport rate in Figure \ref{fig:yutani25-mfr-r} are consistent.\par

To explore the mechanism of mass transport shown in Figure \ref{fig:yutani25-mfr-r}, we first evaluated the $z$-component of gravitational torque received by the gas particle.
\Add{We defined gravitational torque, $\tau_{\rm G,z,i}$, as ${\bf r_{\rm i}}\times m_{\rm i} \ddot {{\bf r}_{\rm i}}$, where $m_{\rm i}$ is gas particle mass.
And, we calculated mass weighted gravitational torque, $\Sigma m_{\rm i} \tau_{\rm G,z,i}/\Sigma m_{\rm i}$, in a 2D cylindarical grid.
Figure \ref{fig:yutani25-grav} shows the gravitational torque distribution.
We can see that negative torques were generated at the front of spiral arms.}
This torque pattern along the spiral arm \Add{is} the expected gravitational torque in a non-axisymmetric disk \citep{lyndenbell1972}.
A similar pattern in observed for M51 is consistent with our results \citep{meidt2013}.\par

%圧力勾配力トルクとの比較
In addition, we calculated torques due to pressure gradient forces, which are defined as follows :
\begin{equation}
	\tau_{\rm prg,z} \equiv \bigg( {\bf r} \times m\frac{{\bf \nabla} {\bf P}}{\rho}\bigg)_z,
\end{equation}
where {\bf r} is the \Add{position vector} from the center of gravity and {\bf P} is the pressure.
In gas systems, pressure gradient forces also have a significant effect on the motion of gas particles. 
Comparing the torques due to the pressure gradient force and self-gravity reveals the force that is truly responsible for the angular momentum transport of the gas.\par

\begin{figure}
	%\epsscale{1.2}
	\includegraphics[width=0.45\textwidth]{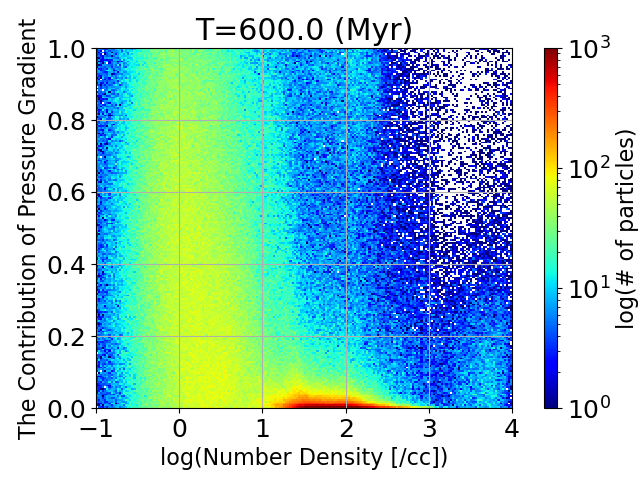}
	%\plotone{Yutani25-torque-ratio.png}
	\caption{The contribution of pressure gradient torque ($|\tau_{\rm prg,z}|/(|\tau_{\rm prg,z}|+|\tau_{\rm G,z}|)$) versus gas number density at 600 Myr in fiducial model.}\label{fig:yutani25-prg}
\end{figure}

Figure \ref{fig:yutani25-prg} shows the contribution from the pressure gradient force torque to the sum of the gravity torque and pressure gradient force torque in \Add{the} fiducial model.
From this figure, it can be seen that the gravity torque is dominant in high-density regions ($n_{\rm H} > 10 {\rm cm^{-3}}$) such as the arms.
In other words, mass transport along the arms is governed by the gravity torque from the arms.\par

\Add{\cite{goldbaum2015, goldbaum2016} also suggested that a mass transport rate of 1 $M_\odot\ {\rm yr^{-1}}$ in Milky Way-like galaxies on galactic disk scales can be explained by gravitational instability with variable gas fractions, and that SN feedback does not play a significant role. 
We also computed the torque from turbulence, and confirmed that gravitational torque is mainly the driving force of the mass flow rate in Appendix \ref{app:stress}.
While our results are consistent with theirs, we further found that mass transport within the central 1 kpc is regulated by the bulge potential. In particular, a stronger bulge potential suppresses non-axisymmetric structures, which in turn reduces radial inflow toward the center (see Figure \ref{fig:yutani25-mfr}).}\par

\subsection{Rapid Clumpy Accretion}\label{sec:clumpy-acc}
\subsubsection{Identification of Gas Clumps}
\begin{figure}
	\includegraphics[width=0.45\textwidth]{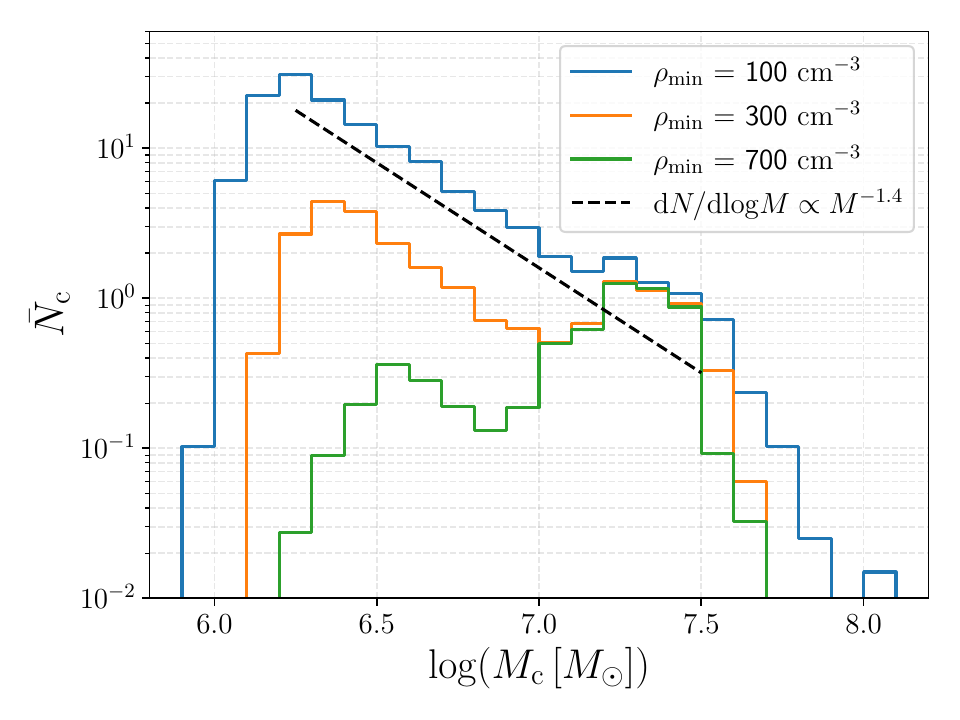}
	\caption{\Add{Mass distribution of gas clumps with varying FoF density thresholds in the fiducial model. 
			$\bar{N}_{\rm c}$ denotes the mean number of clumps per snapshot per logarithmic $x$-bin, averaged over 400 snapshots from 600 Myr to 1 Gyr at a temporal resolution of 1 Myr.}}\label{fig:yutani25-clump-mass}
\end{figure}

\Add{In Section \ref{sec:clumpy-acc}, we focus on clumpy accretion in the fiducial model ($M_{\rm bulge}/M_{\rm disk} = 0.02$).
Following the clump-finding strategy adopted in high-$z$ \Add{disk} galaxy simulation by \cite{ceverino2012}, we first \Add{isolated} gas clumps using a Friends-of-Friends (FoF) algorithm.
To extract dense gas clumps using the FoF algorithm, we constructed tree structures from dense gas particles ($\rho > \rho_{\rm \min}$) and applied a FoF search to the tree nodes.
Figure~\ref{fig:yutani25-clump-mass} shows the gas clump mass spectrum for various density thresholds ($\rho_{\rm \min}$).
From this figure, we found that when the threshold is 100 $\mathrm{cm}^{-3}$, the mass spectrum roughly follows $\mathrm{d}N/\mathrm{d}M \propto M^{-2.4}$ (i.e., $\mathrm{d}N/\mathrm{d}\log M \propto M^{-1.4}$ ) in the mass range $10^6$-$10^{7.5}\ M_\odot$.}\par
\Add{It is known that the mass spectrum of gas clumps in nearby galaxies follows $\mathrm{d}N/\mathrm{d}M \propto M^{-\beta}$ ($\beta\sim2$) \citep[e.g.,][]{engargiola2003, rosolowsky2007, gratier2012, freeman2017, messa2018, andersson2024}.
Our results show a slightly steeper gas-clump mass spectrum than those observed. 
This difference can be attributed to three effects. 
First, in the simulations, we can sample gas clumps without detection biases, whereas observations preferentially detect brighter (more massive) clumps owing to sensitivity and completeness limits. 
Second, the gas surface density in our simulated disk is higher than in typical nearby spiral galaxies, so our catalog probes relatively higher clump masses ($M_{\rm c} \gtrsim 10^6\,M_\odot$) compared to the observational samples. In general, the mass spectrum steepens toward the high-mass end.
Third, the efficiency of SN feedback also slightly affects the result. 
Reducing the SN-feedback efficiency allows low-mass clumps to gain mass and increases the number of high-mass clumps, thereby flattening the mass spectrum (Appendix \ref{app:sn-eff}).}\par

\Add{In this study, we restrict the FoF search to particles with $n_{\rm H} > 700~{\rm cm^{-3}}$ because we are interested in the accretion of dense gas clumps onto the galactic center.
This threshold is well above the characteristic spiral-arm densities ($\sim$ a few $\times\ 10^{2}\ {\rm cm^{-3}}$), allowing us to isolate compact clumps from the surrounding spiral-arm gas even when clumps are embedded within the arms.}

\subsubsection{Lifetimes of Gas Clumps}
\Add{Gas clumps are generally destroyed by SN feedback, but massive clumps can survive it. 
By comparing the self-gravitational potential energy of gas clumps with the energy released by SN explosions, we can estimate the size-mass relation required for a clump to survive for more than 10 Myr.
First, consider simply that 3.3 \% of the gas clump becomes a star because the star formation efficiency $\epsilon_{\rm SF}$ is 0.033 in our models as shown in Equation \ref{eq:sf}.
According to the yields of \cite{nomoto2013}, stars in the 13 $M_\odot$ - 40 $M_\odot$ range are massive enough to undergo core collapse and eventually explode as SNe. 
In addition, assuming a \cite{salpeter1955} IMF over 0.1 $M_\odot$ - 120 $M_\odot$, this corresponds to approximately $3.1 \times 10^{-3}$ core-collapse SNe per solar mass of stars formed.
Therefore, a condition that a gas clump is dynamically stable against the energy feedback of SNe ($E_{\rm ej} = \eta_{\rm SN}\ \times\ 10^{51} $ erg) can be written as
\begin{equation}
	\frac{3}{5}\frac{GM_{\rm c}^2}{R_{\rm c}} > 0.0031 \times \epsilon_{\rm SF} \times\frac{M_{\rm c}}{M_\odot}\times E_{\rm ej},\label{eq:clump-grav-sn}
\end{equation}
where $M_{\rm c}$ is the mass of the gas clump and $R_{\rm c}$ is the size.
Therefore, in the case of $\eta_{\rm SN}$ is unity, we obtain the relationship 
\begin{equation}
	M_{\rm c} \gtrsim 2 \times 10^7 \times \bigg(\frac{R_{\rm c}}{\rm100\ pc}\bigg)\hspace{0.25cm} M_\odot.\label{eq:llclump-mass}
\end{equation}
This relation indicates that more than 2 $\times$ 10$^7$ $M_\odot$ gas clumps would survive for 100 pc scale gas clumps.}\par

\Add{When the mass of a gas clump is lower than $\sim 10^{7}\ M_\odot$, it will be destroyed by SN feedback within $\sim 10$ Myr.
Since the crossing time of a SN shock wave is much shorter than the lifetime of OB stars ($\sim 10$ Myr), the lifetime of such gas clumps is essentially determined by the lifetime of OB-type stars.
The crossing time of a SN shock wave can be estimated as
\begin{equation}
	t_{\rm cross} = \frac{R_{\rm c}}{c_{\rm s}} 
	= \frac{100\ {\rm pc}}{100\ {\rm km\ s^{-1}}} \sim 1\ {\rm Myr},
\end{equation}
where $c_{\rm s}$ is the sound speed of the SN shock wave.}

\begin{figure}
	\includegraphics[width=0.45\textwidth]{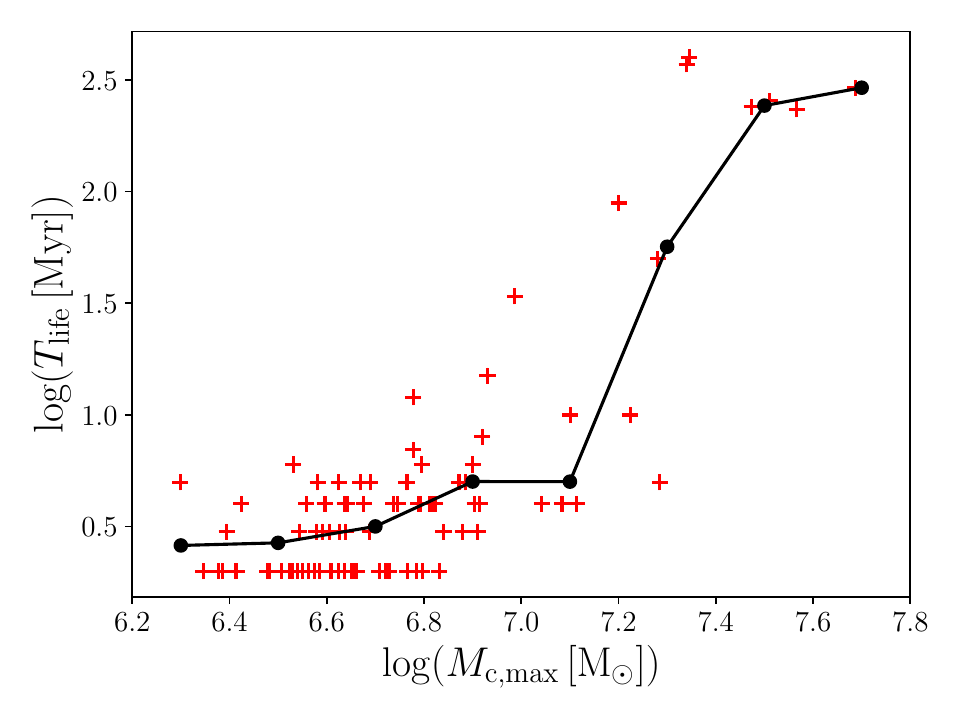}
	\caption{\Add{Final masses and lifetimes of gas clumps identified with a FoF density threshold of 700 ${\rm cm^{-3}}$ in the fiducial model between 600 Myr and 1 Gyr. The black solid line represents the average lifetime of clumps in each gas clump mass bin.}}\label{fig:yutani25-clump-lifetime}
\end{figure}

\Add{By tracking groups of gas particles over time, the lifetime of gas clumps can be estimated as shown in Figure~\ref{fig:yutani25-clump-lifetime}, which plots clump masses against lifetimes.
A gas mass was assumed to have disappeared if it accreted within 600 pc of the galactic center or if it could not be continuously identified.
In our model, long-lived gas clumps (with lifetimes > 10 Myr) tend to be more massive than $10^7\ {\rm M_\odot}$, whereas less massive ($< 10^{7}\ {\rm M_\odot}$) clumps are often disrupted by SN feedback within $\sim$10 Myr.}\par

\subsubsection{Evolution of Long-Lived Gas Clumps}
\begin{figure}
	%\epsscale{1.2}
	%\plotone{Yutani25-clump-evol.png}
	\includegraphics[width=0.45\textwidth]{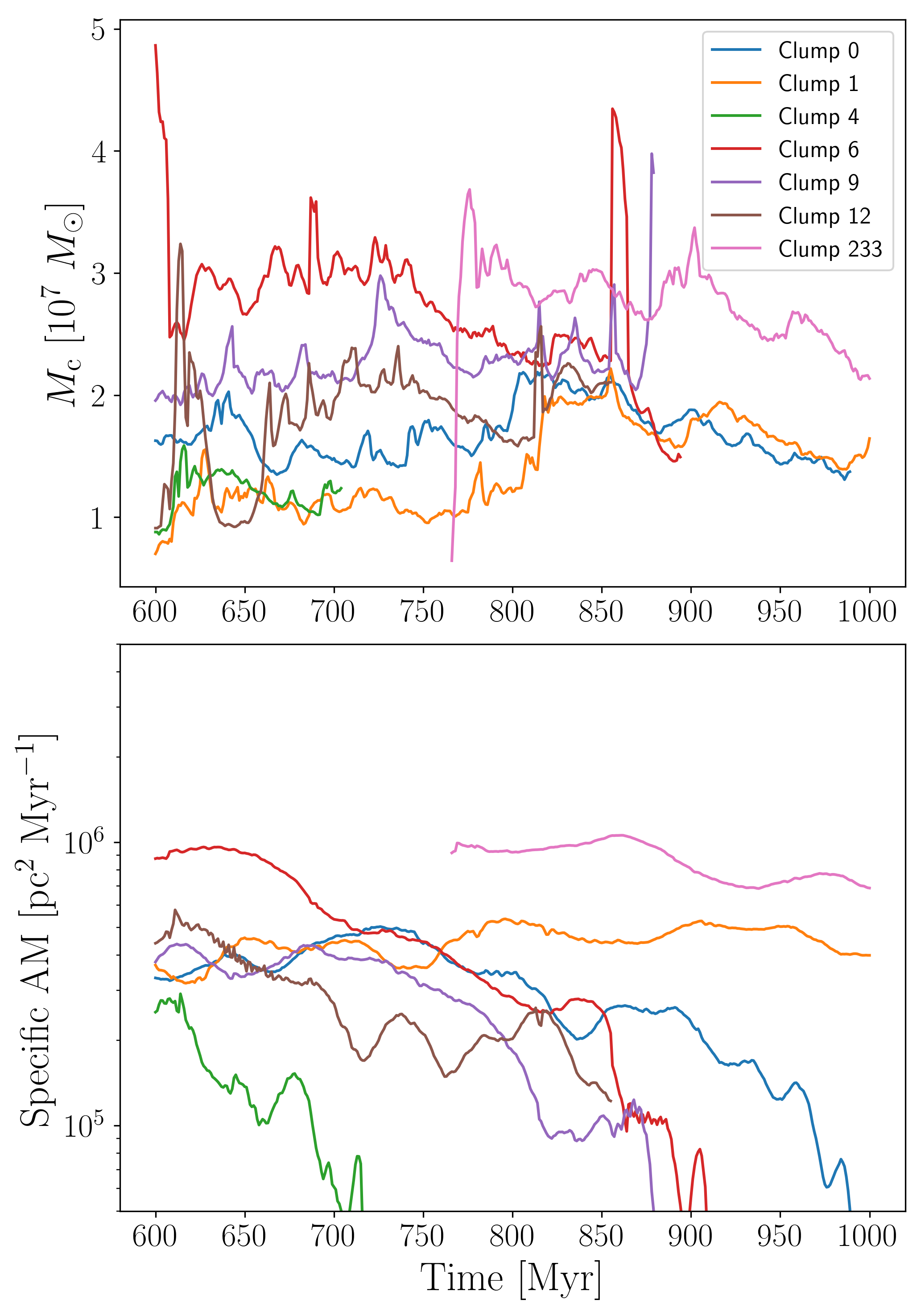}
	\caption{Gas clump mass and specific angular momentum evolution in the fiducial model.}\label{fig:yutani25-clump-evol}
\end{figure}

This study focuses on long-lived clumps above 100 Myr.
The reason for this is that the lifetime of gas clumps falling from the galactic scale ($\sim$ a few kpc) to the CND scale ($\sim$ a few 100 pc) is typically more than 100 Myr.
This is about several times the dynamical time at a few kpc scale.
Figure \ref{fig:yutani25-clump-evol} shows the evolution of gas clump mass and specific angular momentum.
The mass of the gas clumps evolves, holding around $10^7\ M_\odot$.
The mass of individual gas clumps is about 0.1\% of the gas mass of the initial disk. 
Gas clumps gain mass through collisions with arms and other gas clumps and lose mass through collisional and tidal stripping and reduction due to star formation within the gas clump.\par

Figure \ref{fig:yutani25-clump-evol} also shows that the specific angular momentum of the gas clumps decreases rapidly below a few $\times$ 10$^5$ $\rm pc^2\,Myr^{-1}$.
On the other hand, gas clumps with the specific angular momentum higher than a few $\times$ 10$^5$ $\rm pc^2\, Myr^{-1}$ retain their specific angular momentum.
From Figure \ref{fig:yutani25-vc}, it can be seen that the specific angular momentum of a few $\times$ 10$^5$ $\rm pc^2\,Myr^{-1}$ roughly corresponds to a radius of 2 kpc.
This means that gas clumps falling within 2 kpc rapidly decrease their angular momentum and fall further toward a few 100 pc.\par

This rapid clumpy accretion of $10^7$ $M_\odot$ of gas clump occurs on a timescale of 10 Myr, resulting in a mass accretion rate of $\sim$ 1 $M_\odot\ {\rm yr^{-1}}$. 
The accretion events with a timescale of about 10 Myr occur four times during 400 Myr, so the duty cycle is $\sim$ 0.1.
If this duty cycle is taken into account, the mass transport rate on timescales above 100 Myr is around 0.1 $M_\odot\ {\rm yr^{-1}}$.
Note, however, that this duty cycle is highly dependent on when and how many gas clumps are formed.\par

\begin{figure*}
	%\epsscale{1.2}
	%\plotone{Yutani25-clump-orbi.png}
	\includegraphics[width=0.9\textwidth]{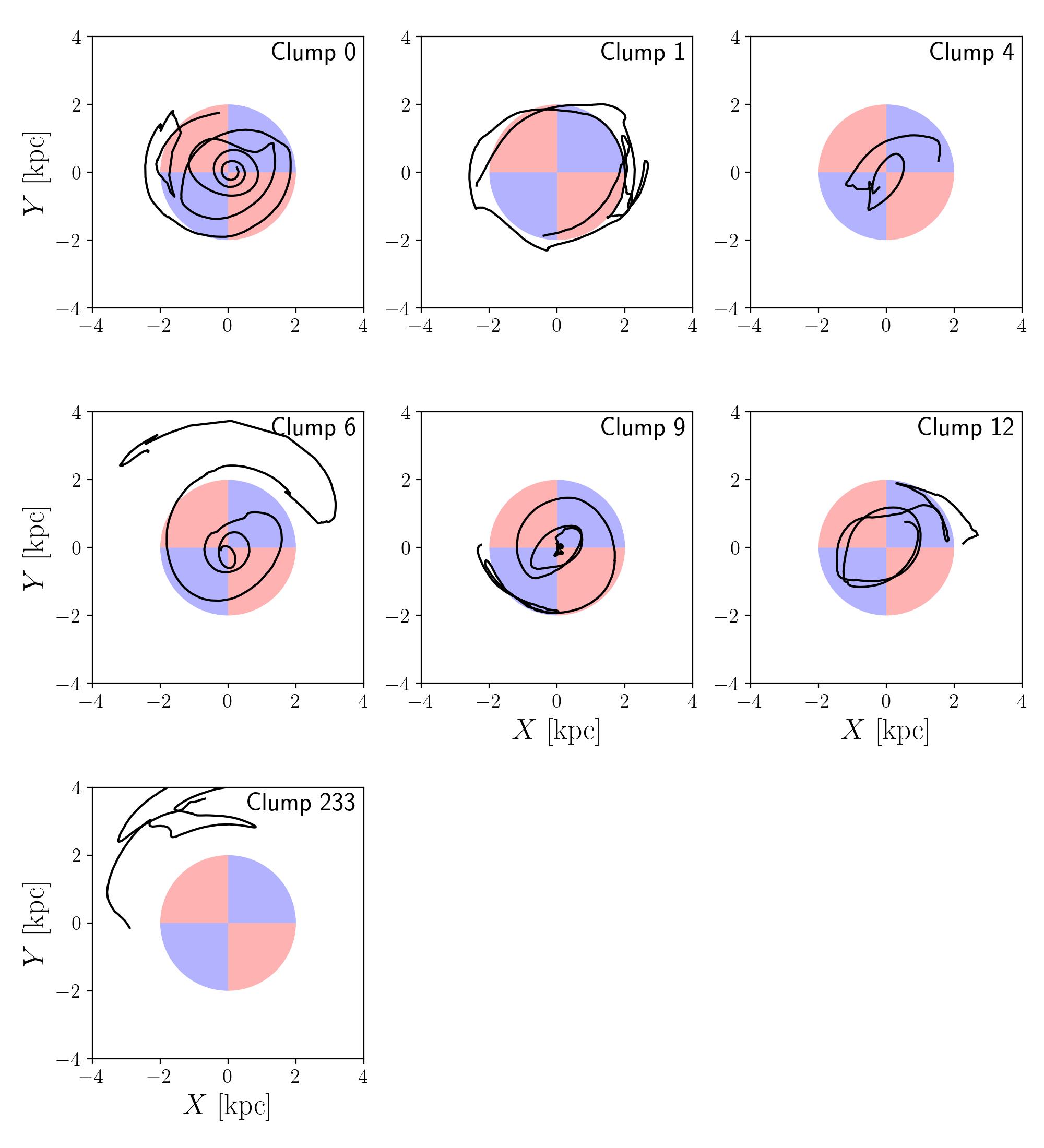}
	\caption{Gas clump orbit in a stationary system with respect to a bar potential fixed to the x-axis in the fiducial model. As the bar potential is fixed to the X-axis and rotates around a semi-clock, the first and third quadrants are the regions subject to negative torque (blue) and the second and fourth quadrants to positive torque (red).}\label{fig:yutani25-clump-orbi}
\end{figure*}

\subsubsection{Clumpy Accretion under the Influence of the Bar}\label{sec:clumpy-acc-damped}
To explore the accretion mechanism of gas clumps, we performed a modal analysis of the stellar disk potential using \AGAMA \citep{vasiliev2019}.
The results show that clumps with delayed elliptical orbits for the bar mode are rapidly accreting to a few 100 pc (Figure \ref{fig:yutani25-clump-orbi}).
This result can be understood by \Add{the} physical nature of the gas orbits in a non-axisymmetric potential. 
\cite{wada1994} analytically showed that the closed orbit of gas in a weak bar potential extends in a direction that precedes the bar potential inside the corotation.
This is a natural consequence of a forced oscillator with dissipation, which corresponds to the epicycle motion of the orbiting material in a galactic potential.
As a result, the gases in these orbits (``damped orbits'') lose their net angular \Add{momentum} by the torque of the bar potential.
Based on this theory, the rapid accretion of the clumps \Add{is} seen in \Add{the} Figures. 
\Add{Figures}~\ref{fig:yutani25-clump-evol} and \ref{fig:yutani25-clump-orbi} can be naturally understood; once a gas clump follows the damped orbit, it loses its orbital angular momentum over one rotational period, its orbit becomes more elongated in the ``negative'' torque domains (the first and third quadrant in Figure \ref{fig:yutani25-clump-orbi}). 
If this process works, the gas clumps in the damped orbits rapidly lose their angular momenta inside the corotation.\par

\begin{figure}
	\epsscale{1.25}
	\centering
	%\plotone{Yutani25-high-mfr-r.pdf}
	\includegraphics[width=0.45\textwidth]{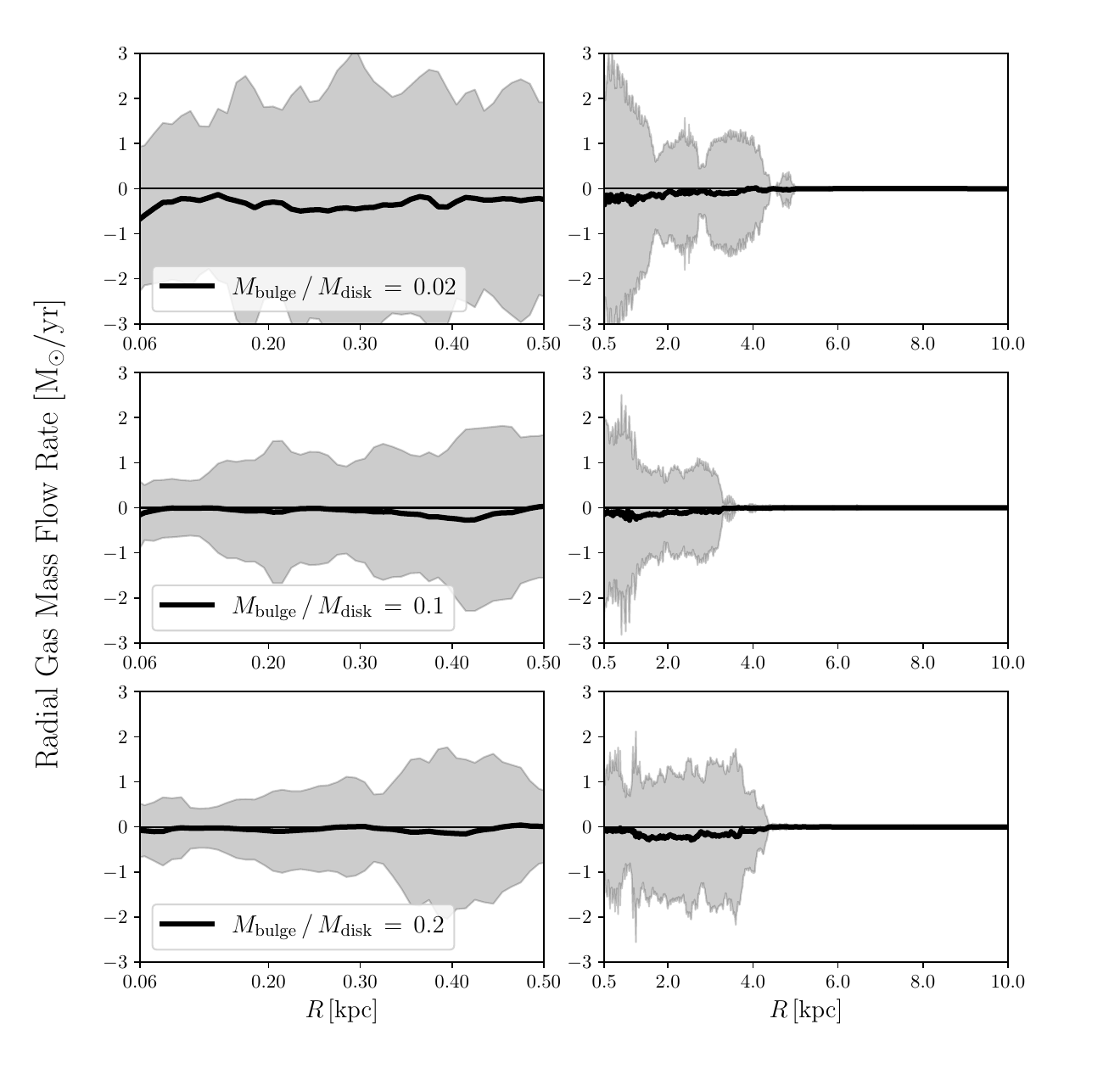}
	\caption{Radial gas mass flow rate in each radius from 60 pc to 10 kpc in three models. This figure is calculated using only high-density particles ($n_{\rm H} > 700$ ${\rm cm^{-3}}$), so this figure focuses only on the dense component of mass transport in Figure \ref{fig:yutani25-mfr-r}. The solid black line shows the averaged gas flow rate\Add{,} and the shaded area shows the standard deviation. To account for the SMBH's motion, mass transport rates from 500 pc to 10 kpc were calculated in a coordinate system stationary with respect to the system's centre of gravity, whereas the rates from 60 pc to 500 pc were calculated in a coordinate system stationary with respect to the SMBH.}\label{fig:yutani25-high-mfr-r}
\end{figure}

Gas clumps falling within the corotational radius of the bisymmetric component of the non-axisymmetric potential (bar mode potential) are rapidly decreasing thier angular momentum because they precede the bar mode potential.
Since the strength of \Add{the} mode 2 potential in the fiducial model varies with time, its corotational radius also varies.
In addition, at a radius of 3 to 4 kpc, where the gas clumps are stationary, it is affected by the potential of a higher mode than mode 2. 
The fate of the gas clumps is determined by nonlinear phenomena: multiple modes, the self-gravity of the gas clumps, and the change in the corotational radius.\par

Figure~\ref{fig:yutani25-high-mfr-r} shows radial high-density gas mass flow rate in three models.
In Figure~\ref{fig:yutani25-mfr-r}, the mass transport rate was evaluated using all particles, while in Figure~\ref{fig:yutani25-high-mfr-r}, the mass transport rate was evaluated using gas particles with $n_{\rm H} > $ 700 ${\rm cm^{-3}}$.
This threshold for high-density gas is the same as the value applied to FoF. 
The curves are averaged over 600 Myr - 1 Gyr.\par

In the fiducial \Add{model,} the mean inflow at \(R\lesssim1\) kpc is about 0.1$M_\odot\, {\rm yr^{-1}}$, fully consistent with the duty-cycle-corrected clump-accretion rate derived above.
By contrast, the other two models exhibit markedly lower high-density inflow at all radii, because their gas orbits do not become elongated.
As a result, when the bulge dominates the central potential, significant rapid clumpy accretion is largely suppressed.\par

\section{discussion}\label{sec:discuss}
\subsection{The condition of rapid clumpy accretion in spiral galaxies}
We found that in galaxies where bar structures and gas clumps coexist, rapid clump accretion via bar potentials occurs episodically.
It is known that the higher the gas fraction, the more likely gas clumps are to form.
However, bar structure is short-lived in \Add{a} gas-rich spiral galaxy.
This is caused by the central mass concentration growth and the bar structure receiving angular momentum from the gas \citep[e.g.,][]{berentzen1998, bournaud2005}.
However, the timescale for the dissolution of the bar structure is a few Gyr, which is sufficiently longer than the cycle of rapid clumpy accretion in our model.
Therefore, once the bar instability occurs in the gas-rich spiral disk, rapid clumpy accretion may well occur. \par

The heavier the spherically symmetric potentials\Add{,} such as the DM halo and bulge relative to the disk mass, the more the bar instability is suppressed\Add{,} and thus the bar structure is harder to develop.
Therefore, rapid clumpy accretion will not occur unless the disk potential is sufficiently dominant relative to the spherically symmetric potential.
In fact, in our heavy bulge model, there are no strong elliptical orbits.\par

\subsection{Relation to Violent Disk instability}
Violent Disk Instability (VDI) arises when a galaxy disk is both gas-rich and highly self-gravitating: 
The high surface density creates local patches where the Toomre parameter drops below unity, causing the \Add{disk} to fragment promptly into giant ($\sim10^{8\text{-}9}\,M_\odot$) clumps \citep{noguchi1999}.
These clumps exchange angular momentum through dynamical friction, lose net angular momentum, and inward within 10 dynamical timescales, or 0.5 Gyr \citep{dekel2009,bournaud2011}.
Their inward migration deposits large reservoirs of gas in the central kpc, thereby fuelling bulge growth and feeding the central SMBH.\par

In disks where VDI occurs, bar structures would not form because the disk would break up violently. 
Sufficient stellar mass is required to form a barred structure. 
Therefore, VDI and rapid clumpy accretion in spiral galaxies are completely different phenomena.
Rapid clumpy accretion is a significant mechanism that can supply mass to the galactic center with high efficiency even in spiral galaxies with modest gas fractions at low redshifts.\par

Interestingly, however, JWST is beginning to discover surprisingly mature massive disk galaxies, some of which already show well-formed spiral structures at $z\sim 2$ \citep[e.g.,][]{wu2023,guo2023}.
If such systems host sufficiently cold, self-gravitating gas, the same rapid clumpy accretion mechanism we \Add{found} here could occur at early epochs as well, resulting in the rapid growth of the SMBH.\par

\section{conclusions}
We calculated isolated spiral galaxies using \textit{N}-body/SPH code \ASURA \citep{saitoh2008,saitoh2013} to explore the mechanism of mass accretion from 10 kpc to 60 pc.
The main results are as follow\Add{s}:
\begin{enumerate}
	\item In our model, there is a steady mass transport process by about 10$^2$ ${\rm cm^{-3}}$ of gas along the arms and a temporary mass transport process by dense gas clumps of about 10$^3$ ${\rm cm^{-3}}$.
	\item A steady mass transport through the spiral arm is caused by the gravitational torque from the arms formed by the stellar disk, which reduces the angular momentum of the gas in front of the arms.
	\item Episodic mass accretion by gas clumps ($> 10^7 M_\odot$) can be rapidly accreted into a few 100 pc by gravitational torque from the bar potential, with a timescale of a few 10 Myr. 
	\item Rapid gas clumpy accretion requires bar potential, which is unlikely to occur in galaxies with dominant spherically symmetric potentials ($M_{\rm bulge}/M_{\rm disk}\ =\ 0.1$ and $0.2$).
	\item On timescales above 100 Myr, steady mass transport by the spiral arm dominates mass transport to the \Add{100 pc from galactic center}, and the mass accretion rate is about 1 $M_\odot$ yr$^{-1}$. On the other hand, on the 10 Myr timescale, when rapid gas clump accretion occurs, the mass accretion rate is also about 1 $M_\odot$ yr$^{-1}$.
\end{enumerate}
Finally, note that the survival of gas clumps may depend on the treatment of stellar feedback.
However, this issue can be addressed only with galaxy-scale simulations that resolve individual gas clouds at sub-parsec scales, fine enough to capture H II regions, and not by relying solely on local molecular-cloud simulations.
The reason is that our calculations show that gas clumps gain mass due to collisions with arms and other gas clumps, and according to \cite{wada2002}, galaxy rotation energy may contribute to the turbulent energy of gas clouds.
For these reasons, sub-pc resolution must be used in galaxy evolution simulations to solve the problem of the survival of gas clumps.

%% IMPORTANT! The old "\acknowledgment" command has be depreciated. It was
%% not robust enough to handle our new dual anonymous review requirements and
%% thus been replaced with the acknowledgment environment. If you try to 
%% compile with \acknowledgment you will get an error print to the screen
%% and in the compiled pdf.
\begin{acknowledgments}
We thank J. Baba and J. Koda for helpful discussions during the development of this research. Numerical simulations were performed by ATERUI III in the Center for Computational Astrophysics (CfCA), National Astronomical Observatory of Japan\Add{,} and Yukawa-21 at the Yukawa Institute Computer Facility. This work was supported by JSPS KAKENHI Grant \Add{Numbers} JP24KJ1833 (NY), JP25H00671 (KW), JP24K07095 (TRS)\Add{,} and JP25H00664 (TRS). \Add{Finaly, we are grateful to the anonymous referee for constructive comments that helped us improve this work.}
\end{acknowledgments}

%% To help institutions obtain information on the effectiveness of their 
%% telescopes the AAS Journals has created a group of keywords for telescope 
%% facilities.
%
%% Following the acknowledgments section, use the following syntax and the
%% \facility{} or \facilities{} macros to list the keywords of facilities used 
%% in the research for the paper. Each keyword is check against the master 
%% list during copy editing. Individual instruments can be provided in 
%% parentheses, after the keyword, but they are not verified.

%\vspace{5mm}
%\facilities{HST(STIS), Swift(XRT and UVOT), AAVSO, CTIO:1.3m,
%CTIO:1.5m,CXO}

%% Similar to \facility{}, there is the optional \software command to allow 
%% authors a place to specify which programs were used during the creation of 
%% the manuscript. Authors should list each code and include either a
%% citation or url to the code inside ()s when available.

\software{\ASURA \citep{saitoh2008, saitoh2013}, \texttt{CELib} \citep{saitoh2017}, \texttt{Phantom-GRAPE} \citep{tanikawa2013}, \AGAMA \citep{vasiliev2019}, \texttt{COLOSSUS} \citep{diemer2018}, \Add{\texttt{PEGASE} \citep{fioc1999}, \texttt{Cloudy} \citep{ferland1998}}}

%% Appendix material should be preceded with a single \appendix command.
%% There should be a \section command for each appendix. Mark appendix
%% subsections with the same markup you use in the main body of the paper.

%% Each Appendix (indicated with \section) will be lettered A, B, C, etc.
%% The equation counter will reset when it encounters the \appendix
%% command and will number appendix equations (A1), (A2), etc. The
%% Figure and Table counter will not reset.

\appendix

\section{Appendix}
\subsection{lower resolution model}\label{app:lower}
We also tested whether rapid clumpy accretion occurs in the lower-resolution model.
Compared to the fiducial model, the gas particle mass is 2.5 times larger, and its gravitational softening increases from 30 pc to 50 pc.
The stellar disk particle mass is likewise multiplied by 2.5, with softening raised from 30 pc to 80 pc.
For the dark matter halo, the particle mass doubles, and softening goes from 50 pc to 100 pc.\par

\begin{figure}
	\centering
	%\plotone{Yutani25-app-clump-evol.png}
	\includegraphics[width=0.45\textwidth]{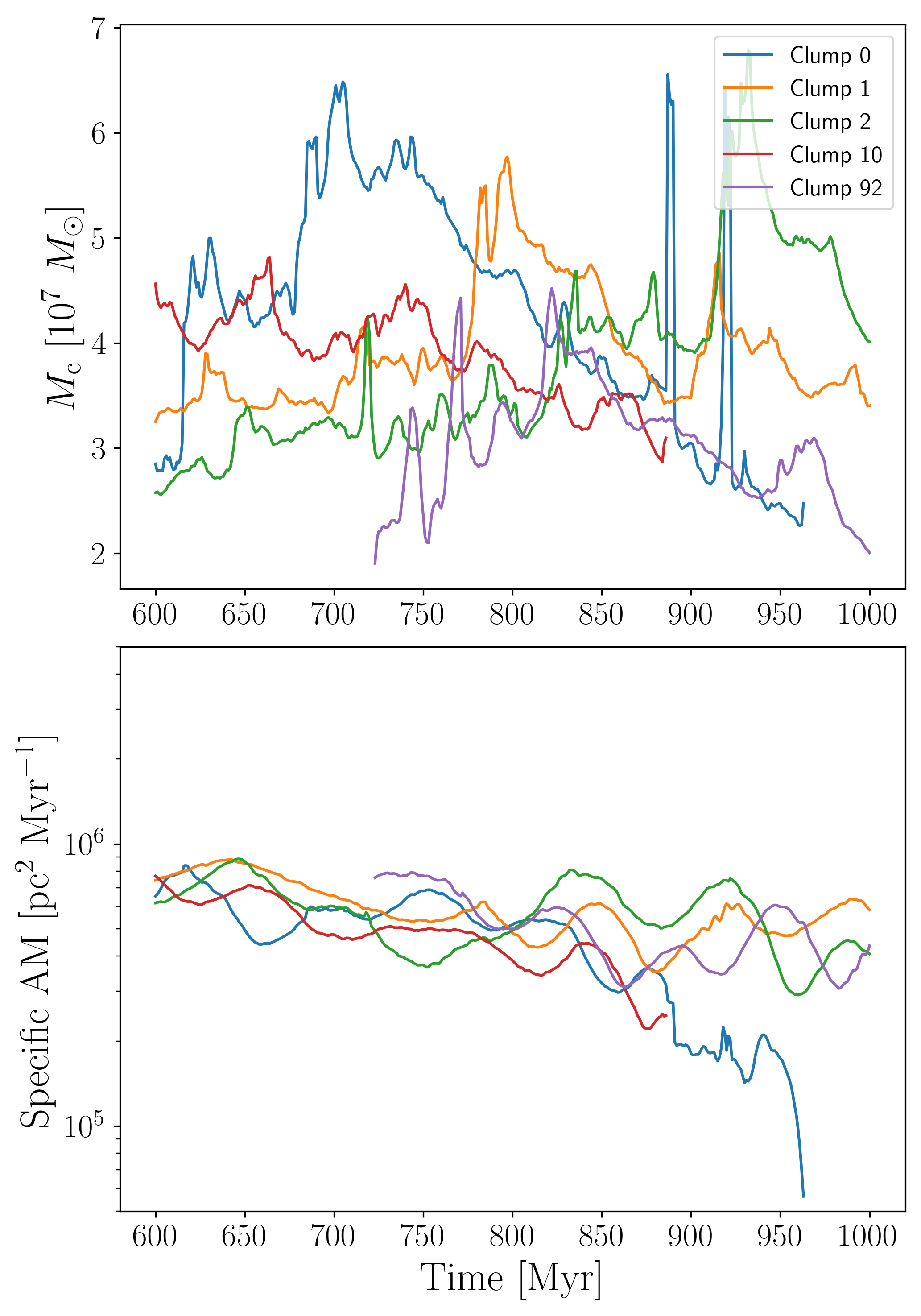}
	\caption{\Add{Evolution of gas clump mass and specific angular momentum in the lower resolution model with $M_{\mathrm{bulge}}/M_{\mathrm{disk}} = 0.02$}. }\label{fig:yutani25-app-clump-evol}
\end{figure}

We confirmed that the gas clump mass does not depend on resolution, and rapid clumpy accretion also occurred in a lower-resolution model (Figure~\ref{fig:yutani25-app-clump-evol}).
Even in the low-resolution model, angular momentum transport occurs on a timescale of 10 Myr, as in the fiducial model, when the specific angular momentum is less than 10$^5$ pc$^2$ Myr$^{-1}$.
The two instantaneous mass increases between 850 Myr and 950 Myr for Clump ID 0 are due to collisions with Clump ID 10.
However, Clump ID 10 collides and splits with Clump ID 0 at 890 Myr and is recognized as a different gas clump in the clump identification routine.\par

\begin{figure}
	\centering
	%\plotone{Yutani25-app-clump-orbi.png}
	\includegraphics[width=0.45\textwidth]{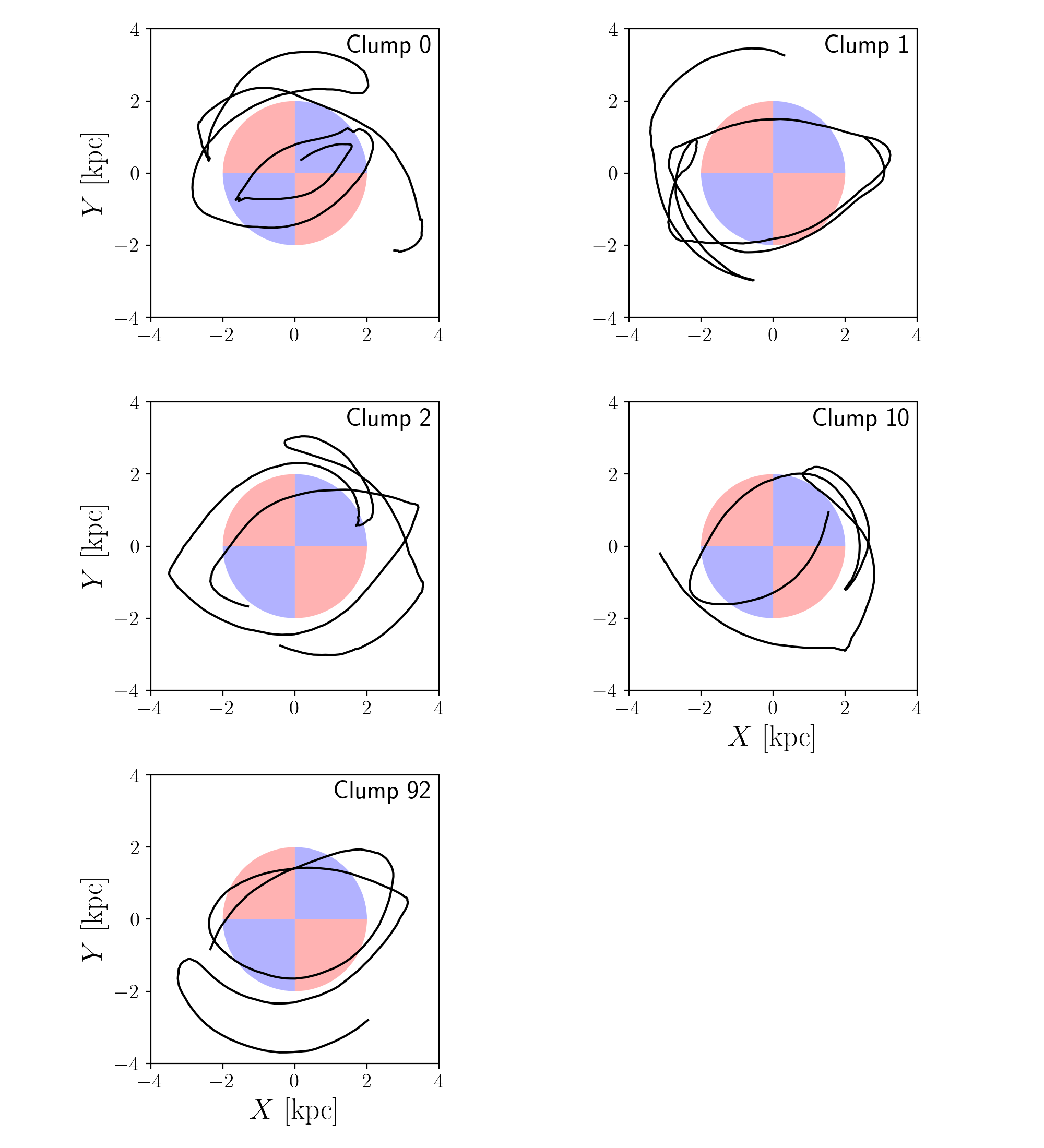}
	\caption{\Add{Orbits of gas clumps in the frame where the bar potential is held fixed and aligned with the x-axis, for the low resolution model with $M_{\mathrm{bulge}}/M_{\mathrm{disk}} = 0.02$}. As the bar potential is fixed to the X-axis and rotates around a \Add{counterclockwise}, the first and third quadrants are the regions subject to negative torque (blue) and the second and fourth quadrants to positive torque (red).}\label{fig:yutani25-app-clump-orbi}
\end{figure}

The orbit of the rapidly accreting gas clump is plotted in Figure~\ref{fig:yutani25-app-clump-orbi}.
From this figure, it can be seen that the accretion is occurring.
Thus, rapid clumpy accretion by bar mode potential is confirmed even by the low-resolution model.\par

\subsection{Sensitivity of gas clump survival to SF efficiency}\label{app:sf-eff}
\begin{figure}
	\centering
	\includegraphics[width=0.95\textwidth]{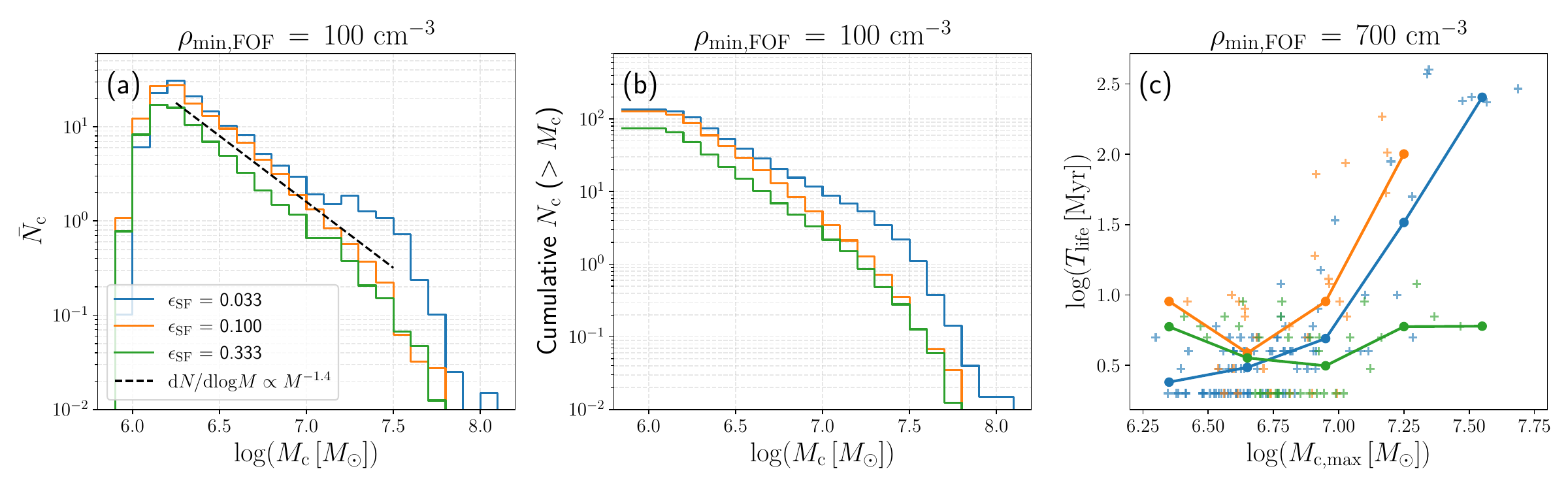}
	\caption{\Add{
			Mass distribution and lifetime of gas clumps for the fiducial model ($M_{\mathrm{bulge}}/M_{\mathrm{disk}}=0.02$), showing how they vary with $\epsilon_{\rm SF}$ (three values: 0.0333 (fiducial), 0.100 and 0.333) as defined in equation~(\ref{eq:sf}).
			The blue, orange, and green colors denote $\epsilon_{\rm SF}$ = 0.033, 0.100, and 0.333, respectively.
			(a) Mass distribution of gas clumps. $\bar{N}_{\rm c}$ denotes the mean number of clumps per snapshot per logarithmic x-bin, averaged over 400 snapshots from 600 Myr to 1 Gyr at a temporal resolution of 1 Myr. Clumps are identified by applying a FoF grouping to tree nodes constructed from gas with $n_{\rm H} \ge 100 \mathrm{cm^{-3}}$
			(b) Cumulative clump mass spectrum derived from the upper panel. 
			(c) Lifetime-mass relation of gas clumps. The x-axis shows the maximum mass during the lifetime of the gas mass. The solid lines represent the average lifetime of clumps in each gas clump mass bin. Clumps are identified by applying a FoF grouping to tree nodes constructed from gas with $n_{\rm H} \ge 700 \mathrm{cm^{-3}}$ in panel (c).}}\label{fig:yutani25-app-clump-sf}
\end{figure}

\Add{We investigate the impact of SF efficiency in equation (\ref{eq:sf}) on the survival of long-lived gas clumps in our simulations.
Figure~\ref{fig:yutani25-app-clump-sf} shows the gas clump mass spectrum and lifetime with different star formation efficiency, $\epsilon_{\rm SF}$, for the fiducial model.
In the text, we use $\epsilon_{\rm SF}$ = 0.033 based on \cite{saitoh2008}.
In Panels (a) and (b), the FoF gas density threshold is set to 100 $\mathrm{cm^{-3}}$ in order to obtain an unbiased mass spectrum over a wide range.
In contrast, Panel (c) uses a threshold of 700 $\mathrm{cm^{-3}}$.
This high density is necessary, even when gas clumps are located with spiral arms, detection requires a density threshold greater than the typical gas arm density ($\sim$100 $\mathrm{cm^{-3}}$).}\par

\Add{Figure \ref{fig:yutani25-app-clump-sf} shows that increasing the star formation efficiency leads to a relative decrease in high-density gas clumps. In particular, the model with $\epsilon = 0.33$ exhibits a clear reduction in the population of low-mass clumps. 
As a result, the number of long-lived clumps with $M_{\rm c} > 10^7\ M_\odot$ decreases by approximately 60\% when $\epsilon$ is increased from 0.033 to 0.33. 
This trend is consistent with the findings of \cite{saitoh2008}, which reported that higher star formation efficiency suppresses the formation of dense gas. 
Nevertheless, even in the $\epsilon_{\rm SF} = 0.33$ model, clumps with $M_{\rm c} > 10^7\ M_\odot$ are still present.
However, in the $\epsilon_{\rm SF} = 0.33$, there are no long-lived gas clumps ($T_{\rm life} > 100$ Myr).
This is because higher star-formation efficiency leads to stronger SN feedback, and gas is consumed more rapidly by star formation.}\par

\subsection{Sensitivity of gas clump survival to SN feedback efficiency}\label{app:sn-eff}
\begin{figure}
	\centering
	\includegraphics[width=0.95\textwidth]{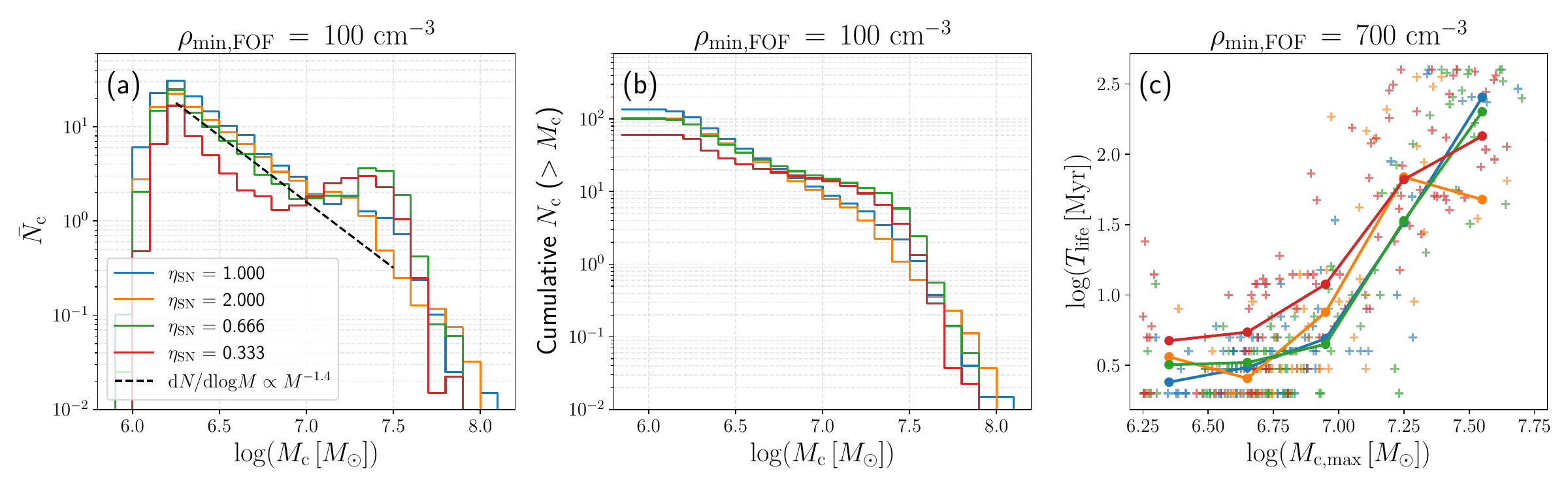}
	\caption{\Add{
			Mass distribution and lifetime of gas clumps for the fiducial model ($M_{\mathrm{bulge}}/M_{\mathrm{disk}}=0.02$), showing how they vary with $\eta_{\rm SN}$ (four values: 1.000 (fiducial), 2.000, 0.666 and 0.333).
			The blue, orange, green, and red colors denote $\eta_{\rm SN}$ = 1.0, 2.0, 0.666, and 0.333, respectively.
			(a) Mass distribution of gas clumps. $\bar{N}_{\rm c}$ denotes the mean number of clumps per snapshot per logarithmic x-bin, averaged over 400 snapshots from 600 Myr to 1 Gyr at a temporal resolution of 1 Myr. Clumps are identified by applying a FoF grouping to tree nodes constructed from gas with $n_{\rm H} \ge 100\ \mathrm{cm^{-3}}$
			(b) Cumulative clump mass spectrum derived from the upper panel. 
			(c) Lifetime-mass relation of gas clumps. The x-axis shows the maximum mass during the lifetime of the gas mass. The solid lines represent the average lifetime of clumps in each gas clump mass bin. Clumps are identified by applying a FoF grouping to tree nodes constructed from gas with $n_{\rm H} \ge 700 \mathrm{cm^{-3}}$ in panel (c).}}\label{fig:yutani25-app-clump-sn}
\end{figure}

\Add{It is known that gas clump survival depends on treatment of stellar feedback \citep{mayer2016, andersson2024}.	
We also investigate the relation of survival of gas clumps and SN feedback efficiency.
As shown in Section \ref{sec:numerical}, single SN feedback energy is defined as $E_{\rm ej} = \eta_{\rm SN} \times 10^{51} {\rm erg}$.
In other words, the higher the SN feedback efficiency ($\eta_{\rm SN}$) imparts more energy to the surrounding gas.
Figure~\ref{fig:yutani25-app-clump-sn} shows the mass spectrum and lifetimes of gas clumps for different values of $\eta_{\rm SN}$.
In the text, we use $\eta_{\rm SN}$ = 1.0 based on \cite{hopkins2018}.
The threshold densities are the same as those in Figure \ref{fig:yutani25-app-clump-sf} for panels (a), (b), and (c) in Figure \ref{fig:yutani25-app-clump-sn}.
In Panels (a) and (b), the FoF gas density threshold is set to 100 $\mathrm{cm^{-3}}$ in order to obtain an unbiased mass spectrum over a wide range.}

\Add{Panel~(a) of Figure~\ref{fig:yutani25-app-clump-sn} shows that the mass spectrum becomes flatter when the SN-feedback efficiency decreases from $\eta_{\rm SN}$ = 2.0 to 0.333. This is because small gas clumps grow into larger ones. \cite{andersson2024} also found a flatter spectrum when SNe feedback was not included. Gas clumps above $10^7$ $M_\odot$ appear in all cases. However, clumps above $10^9$ $M_\odot$ do not appear. This is consistent with \cite{mayer2016}, who simulated galaxy disks with both blast-wave feedback and their new feedback model. In both cases, no gas clumps larger than $10^9$ $M_\odot$ were found.}\par

\subsection{Reynolds Stress versus Gravitational Stress}\label{app:stress}
\Add{We also examined the impact of turbulence on angular momentum transport. Following \cite{lodato2005}, we compute and azimuthally integrate the Reynolds and gravitational stresses as
\begin{equation}
	T_{\rm r\phi,\,Reynolds}(r)\ =\ \int_{0}^{2\pi}\Sigma(r, \phi)\,v_{\rm r}(r, \phi)\,(v_{\phi}(r, \phi)\ -\ \overline{v}_{\phi}(r))\,d\phi,
\end{equation}
and
\begin{equation}
	T_{\rm r\phi,\,Gravity}(r)
	\ =\ \int_{0}^{2\pi}\int_{-100,\mathrm{pc}}^{100\,\mathrm{pc}}
	\frac{g_{r}(r,\phi,z)\,g_{\phi}(r,\phi,z)}{4\pi G}\,dz\,d\phi,
\end{equation}
where $\overline{v}_{\phi}(r)$ is the surface density weighted azimuthal average at each radius.}

\begin{figure}
	\centering
	\includegraphics[width=0.5\linewidth]{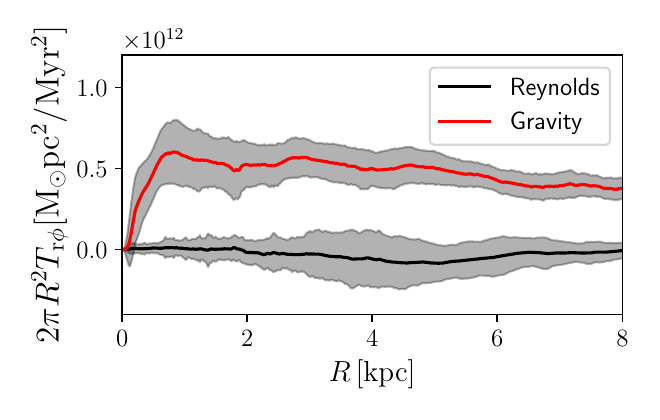}
	\caption{\Add{The radial profile of the Reynolds stress and the gravitational stress in our fiducial model from 600 Myr to 1 Gyr. The solid curves are averaged radial profiles of 400 snapshots from 600 Myr to 1 Gyr. The shaded region shows standard deviation of each stress.}}\label{fig:stress}
	\label{fig:yutani25-app-stress}
\end{figure}

\Add{Figure \ref{fig:yutani25-app-stress} shows the radial profile of the torque from gravitational and Reynolds stresses in the fiducial model.
To compute this radial profile, we used 400 snapshots from 600 Myr to 1 Gyr. The gravitational stress is sufficiently larger than the Reynolds stress. The gravitational stress is about 5 times larger than Reynolds stress.}

\begin{figure}
	\centering
	\includegraphics[width=0.5\linewidth]{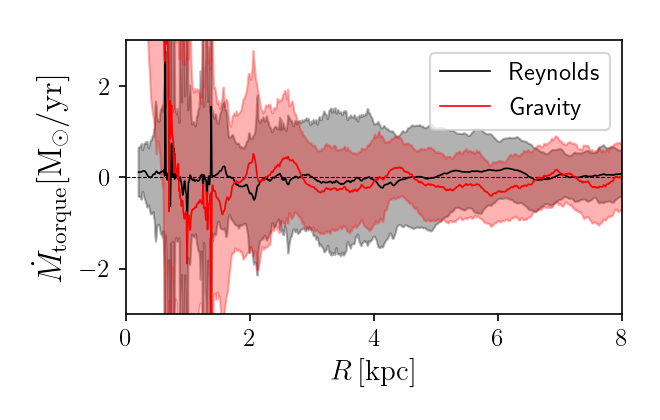}
	\caption{\Add{The radial profile of mass flow rate under assumption of steady states and mass conservation. The solid curves are averaged radial profiles of 400 snapshots from 600 Myr to 1 Gyr. The shaded region shows standard deviation of each mass flow rate as same as Figure \ref{fig:yutani25-mfr-r}.}}
	\label{fig:yutani25-app-mflux}
\end{figure}

\Add{Referring to equation (103) of \cite{armitage2022}, we also derived the relationship between the mass transport rate and the torque from Navier Stokes equation and continuity equation, and evaluated the mass accretion rate as
\begin{equation}
	\dot{M}_{\rm torque} = \Sigma r \overline{v}_r = \bigg(\frac{d rv_{\rm \phi}}{dr}\bigg)^{-1}\frac{d}{dr}\int_0^{2\pi} r^2 T_{\rm r\phi}(r, \phi) d\phi.
\end{equation}}

\Add{This equations based on assumption of steady states and mass conservation. So, the gas consumed by star formation is not considered. 
However, as shown in Figure \ref{fig:yutani25-app-mflux}, it is possible to compare the mass flow rates driven by each torque.
In Figure \ref{fig:yutani25-app-mflux}, we confirm that the gravitational torque is mainly the driving force of the mass flow rate in our model as confirmed in \cite{goldbaum2015, goldbaum2016}.}

%% For this sample we use BibTeX plus aasjournals.bst to generate the
%% the bibliography. The sample631.bib file was populated from ADS. To
%% get the citations to show in the compiled file do the following:
%%
%% pdflatex sample631.tex
%% bibtext sample631
%% pdflatex sample631.tex
%% pdflatex sample631.tex

\bibliography{yutani25}{}
\bibliographystyle{aasjournal}

%% This command is needed to show the entire author+affiliation list when
%% the collaboration and author truncation commands are used.  It has to
%% go at the end of the manuscript.
%\allauthors

%% Include this line if you are using the \added, \replaced, \deleted
%% commands to see a summary list of all changes at the end of the article.
%\listofchanges

\end{document}